\documentclass[aps,prx,twocolumn,superscriptaddress,floatfix]{revtex4-2}

\usepackage{graphicx}
\usepackage{dcolumn}

\usepackage{bm}
\usepackage{amssymb}
\usepackage{amsmath}
\usepackage{microtype}
\usepackage{xfrac}
\usepackage{array}
\usepackage[dvipsnames]{xcolor}
\usepackage[colorlinks,plainpages=false,linkcolor=blue,urlcolor=blue,citecolor=blue,pdfpagemode=UseNone,pdfstartview=FitBH]{hyperref}
\usepackage{nicematrix}

\newcommand{\angstrom}{\text{\normalfont\AA}}

\begin{document}

\title{Proximate Spin Liquid Ground State Arising from Competing Stripy and 120\texorpdfstring{$^\circ$}~ Spin Correlations in the Triangular Quantum Antiferromagnet \texorpdfstring{ErMgGaO$_4$}~}

\author{S.~H.-Y.~Huang}
\affiliation{Department of Physics and Astronomy, McMaster University, Hamilton, Ontario L8S 4M1, Canada}

\author{S.~Petit}
\affiliation{Laboratoire L\'eon Brillouin, CEA, CNRS, Université Paris-Saclay, France}

\author{B.~Yuan}
\affiliation{Department of Physics and Astronomy, McMaster University, Hamilton, Ontario L8S 4M1, Canada}

\author{Z.~W.~Cronkwright}
\affiliation{Department of Physics and Astronomy, McMaster University, Hamilton, Ontario L8S 4M1, Canada}

\author{C. Pinvidic}
\affiliation{Department of Physics and Astronomy, McMaster University, Hamilton, Ontario L8S 4M1, Canada}
\affiliation{Ecole Normale Superiere de Lyon, Lyon, France}

\author{Y. Wang}
\affiliation{Department of Physics and Astronomy, McMaster University, Hamilton, Ontario L8S 4M1, Canada}

\author{E.~M.~Smith}
\affiliation{Department of Physics and Astronomy, McMaster University, Hamilton, Ontario L8S 4M1, Canada}

\author{S.~Bhattacharya}
\affiliation{Universit\'e Paris-Saclay, CNRS, Laboratoire de Physique des Solides, 91405 Orsay, France}

\author{C.~Yang}
\affiliation{Department of Chemistry, Princeton University, USA}

\author{J.-M.~Zanotti}
\affiliation{Institut Laue Langevin, Grenoble, France}
\affiliation{Laboratoire L\'eon Brillouin, CEA, CNRS, Université Paris-Saclay, France}

\author{Q.~Berrod}
\affiliation{Institut Laue Langevin, Grenoble, France}

\author{M.~B.~Stone}
\affiliation{Neutron Scattering Division, Oak Ridge National Laboratory, Oak Ridge, Tennessee 37831, USA}

\author{A.~I.~Kolesnikov}
\affiliation{Neutron Scattering Division, Oak Ridge National Laboratory, Oak Ridge, Tennessee 37831, USA}

\author{R.J.~Cava}
\affiliation{Department of Chemistry, Princeton University, USA}

\author{E.~Kermarrec}
\affiliation{Universit\'e Paris-Saclay, CNRS, Laboratoire de Physique des Solides, 91405 Orsay, France}

\author{B.~D.~Gaulin}
\affiliation{Department of Physics and Astronomy, McMaster University, Hamilton, Ontario L8S 4M1, Canada}
\affiliation{Universit\'e Paris-Saclay, CNRS, Laboratoire de Physique des Solides, 91405 Orsay, France}
\affiliation{Brockhouse Institute for Materials Research, McMaster University, Hamilton, Ontario L8S 4M1, Canada}
\affiliation{Canadian Institute for Advanced Research, 661 University Avenue, Toronto, Ontario M5G 1M1, Canada.}

\date{\today}

\begin{abstract} 
ErMgGaO$_4$ is a quantum antiferromagnet wherein the pseudospin-1/2 degrees of freedom of Er$^{3+}$ decorate two-dimensional triangular planes separated by disordered non-magnetic bilayers of Mg$^{2+}$ and Ga$^{3+}$.  Its sister compound, YbMgGaO$_4$, has attracted much interest as a quantum spin liquid ground state candidate, although the presence of the disordered Mg-Ga bilayers adds complexity to this description.  In contrast, our powder ErMgGaO$_4$ sample shows a clear spin glass transition near $T_g \sim 2.5$~K, about 1/6 of its Curie-Weiss temperature.  We have carried out new inelastic neutron scattering measurements on these powder ErMgGaO$_4$ samples.  At high energies, these show the expected crystalline electric field (CEF) transitions within the $J=15/2$ multiplet of Er$^{3+}$, but with the first excited CEF level sufficiently low in energy ($\sim$ 3~meV) so as to allow the possibility that virtual CEF transitions influence the exchange couplings.  At zero energy transfer, we observe diffuse elastic scattering which is analysed using Warren lineshapes appropriate for two dimensional correlations.  This reveals dominant 2D stripy correlations below $T_g$, coexisting with 2D 120$^\circ$-type correlations that persist above $T_g$.   At low temperatures, the low energy inelastic component of the scattering shows a continuum with bandwidth of $\sim$ 0.8~meV. This dynamic magnetic spectral weight can be modeled at all $Q$, energies, and temperatures as the sum of high energy and low energy damped harmonic oscillators (DHO), with the high energy DHO defining the bandwidth of $\sim$ 0.8~meV. We use linear spin wave theory to model this inelastic scattering and to estimate its spin Hamiltonian parameters in terms of a $J_1-J_2-\Delta$ model on the triangular lattice.  This gives a good description of the low lying spectral weight for ErMgGaO4, and allows us to place it on the theoretical $J_1-J_2-\Delta$ phase diagram with $\frac{J_1}{J_2}=0.13 \pm 0.03$ and $\Delta=0.4 \pm 0.1$, which is close to the expected quantum phase boundary between the spin liquid and the stripy ordered phases.
\end{abstract}

\maketitle

\section{Introduction}

Triangular lattice antiferromagnets were among the first model systems for which geometrical frustration was investigated~\cite{Collins1997, Balentsintro2010}. Pioneering theoretical work by Wannier clearly showed that the ground state for classical Ising antiferromagnetism on the triangular lattice was disordered~\cite{Wannier1950}, while Anderson considered a quantum case with antiferromagnetically-coupled $S=1/2$ degrees of freedom on the triangular lattice for the first potential realization of a resonating valence bond state~\cite{Anderson1973}.

More recently, magnets based on pseudospin-1/2 degrees of freedom occupying a stacked triangular lattice have attracted much attention~\cite{Li2015, Xu2016, Li2017, Paddison2017, Fritsch2017, Chernyshev2017, Liu2018, Cava2018, Kojima2018, Shu2019, Bordelon2019, Gegenwart2019, Cai2020, Haravifard2021, Steinhardt2021, Kulbakov2021, Schmidt2021, Steinhardt2021}.  This is especially so for those based on Yb$^{3+}$ wherein an effective spin-1/2 crystalline electric field (CEF) doublet is typically separated by a large energy gap from the excited CEF states, and the resulting pseudospins with effective spin-1/2 symmetry interact via anisotropic exchange at low temperature~\cite{Li2015, Xu2016, Li2017, Paddison2017, Chernyshev2018, Liu2018, Bordelon2019, Gegenwart2019, Haravifard2021, Steinhardt2021}. While there are several such families of Yb$^{3+}$ triangular magnets, attention originally focused on YbMgGaO$_4$ ~\cite{Li2015, Xu2016, Li2017, Paddison2017, Chernyshev2017, Gegenwart2019, Haravifard2021, Steinhardt2021} as a potential candidate for such a quantum spin liquid (QSL) ground state ~\cite{Balentsintro2010}.  This triangular antiferromagnet has a LuFe$_2$O$_4$-type structure in the $R\bar{3}m$ space group, with triangular layers of Yb$^{3+}$ rare-earth ions interleaved with disordered triangular bilayers of non-magnetic Mg$^{2+}$ and Ga$^{3+}$ ions. The disorder within the Mg-Ga bilayers can induce weak positional disorder by displacing the rare-earth ion away from the otherwise perfect triangular layer to create two rare-earth positions (one just above and one just below the triangular layer) with a random 50$\%$ occupancy~\cite{Li2015, Cava2018}. The displacement between these two rare earth positions is minimal~\cite{Li2015, Cava2018}, and we will henceforth consider the rare-earth sublattice as being a flat network of equilateral triangles. Nonetheless, this weak disorder, and the much stronger disorder associated with the approximate random mixing of Mg$^{2+}$ and Ga$^{3+}$ on the bilayers that interleave with the rare earth layers along the stacking direction~\cite{Li2015, Paddison2017, Cava2018}, significantly complicate the description of quantum triangular antiferromagnets such as the previously studied YbMgGaO$_4$~\cite{Chernyshev2017}, and its sister quantum magnet ErMgGaO$_4$, which we report here. 

YbMgGaO$_4$ was originally proposed as a QSL ground state material, despite the aforementioned disorder, as it displayed many of the hallmarks of such a quantum disordered state, including the absence of long range order or spin freezing, and a continuum of magnetic excitations at low temperatures reminiscent of a spinon continuum~\cite{Paddison2017}.  More recently a spin glass freezing anomaly was found at very low temperatures~\cite{Li2015, Paddison2017, Li2017}, and a theoretical proposal for disorder-induced QSL ``mimicry" was introduced~\cite{Chernyshev2017}. These developments tended to shift the discussion of YbMgGaO$_4$'s ground state towards the question: ``Would YbMgGaO$_4$ display a QSL ground state in the absence of disorder?"  This in turn has led to a greater understanding of the ground state of a quantum triangular lattice magnet with anisotropic exchange, and indeed this overall phase diagram has turned out to be very rich and to host QSL ground states~\cite{Chernyshev2017, Chernyshev2018, Chernyshev2019, Steinhardt2021}. 
            
\section{Outline}

In this paper we report neutron scattering and magnetic susceptibility measurements on a new, near phase-pure, powder sample of ErMgGaO$_4$. Curie-Weiss analysis of the magnetic susceptibility is approximately consistent with earlier work~\cite{Steinhardt2021} and gives an antiferromagnetic Curie-Weiss temperature of $\Theta_{\mathrm{CW}} = -14$~K.  At lower temperatures a bifurcation between the field-cooled (FC) and the zero-field-cooled (ZFC) susceptibility is observed, which we interpret as a spin glass freezing transition near $T_g \sim 2.5$~K.

We report high-energy inelastic neutron scattering spectra measured from powder ErMgGaO$_4$ and we use the energies and intensities of the measured CEF excitations to refine the CEF Hamiltonian appropriate for Er$^{3+}$ in ErMgGaO$_4$. Surprisingly, we find a very low energy first-excited CEF state near 3~meV. Van Vleck susceptibility analysis allows us to account for the high temperature susceptibility, and high field magnetization, in our powder samples.

We also report low-energy inelastic neutron scattering measurements and use the measured magnetic signal to investigate the static spin correlations and collective spin dynamics of this system. We report a continuous distribution of magnetic spectral weight at low $Q$, with a bandwidth of $\sim$ 0.8~meV. Analysis using linear spin wave theory (LSWT) carried out using the SpinW\citep{Toth2015} and ``SpinWave"~\cite{PetitDamay} packages, can account for the bandwidth and broad $Q$-dependence of the magnetic dynamic spectral weight.  This analysis allows us to approximately place ErMgGaO$_4$ within the theoretical $J_1$-$J_2$-$\Delta$ phase diagram for triangular antiferromagnets at zero temperature, with $\frac{J_1}{J_2}=0.13 \pm0.03$ and $\Delta=0.4 \pm 0.1$. Within this model, $J_1$ is the near-neighbour and $J_2$ is the next-near-neighbour exchange strength, while $\Delta$ characterizes the anisotropy of each exchange. This places ErMgGaO$_4$ within the ordered stripy N\'eel phase, but close to its quantum phase boundary with a spin liquid phase.  

Phenomenologically, the full inelastic magnetic spectrum of ErMgGaO$_4$ is shown to be successfully modeled using the sum of two damped harmonic oscillators, allowing its temperature dependence to be studied. 

Finally we present analysis of the strong, elastic diffuse scattering that develops below $T_g$~$\sim$~2.5~K in ErMgGaO$_4$.  The $Q$-dependence of the elastic neutron scattering is analysed using two Warren lineshapes, that describe the two dimensional frozen spin correlations corresponding to a non-collinear 120$^\circ$ N\'eel state and a stripy N\'eel state. This analysis shows that stripy N\'eel correlations ultimately dominate below $T_g$~$\sim$~2.5~K, but these vanish above $T_g$, leaving only the non-collinear 120$^\circ$ N\'eel correlations. 

\begin{figure}[]
\centering
  \includegraphics[width=0.48 \textwidth]{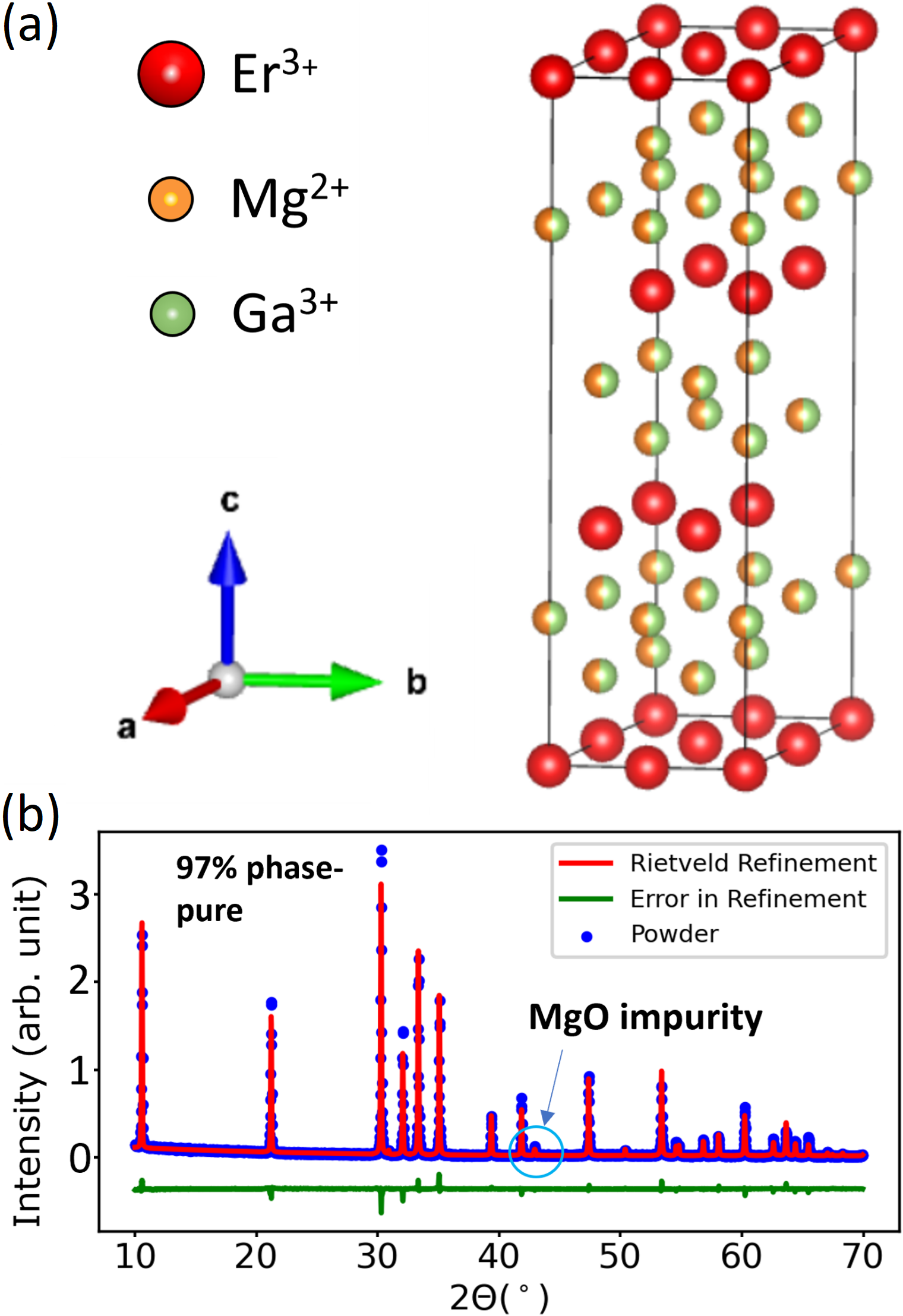}
\caption{(a)~The crystal structure of ErMgGaO$_4$, showing the triangular planes of Er$^{3+}$ interleaved with disordered triangular bilayers of Mg$^{2+}$ and Ga$^{3+}$. (b)~The x-ray diffraction data measured from our powder sample of ErMgGaO$_4$, along with our Rietveld refinement to this data. The sample is 97$\%$ phase pure with the main impurity being unreacted MgO at the $\sim$ 2$\%$ level.}
   \label{Figure1} 
\end{figure}  

\begin{figure}[]
\centering
  \includegraphics[width=0.48 \textwidth]{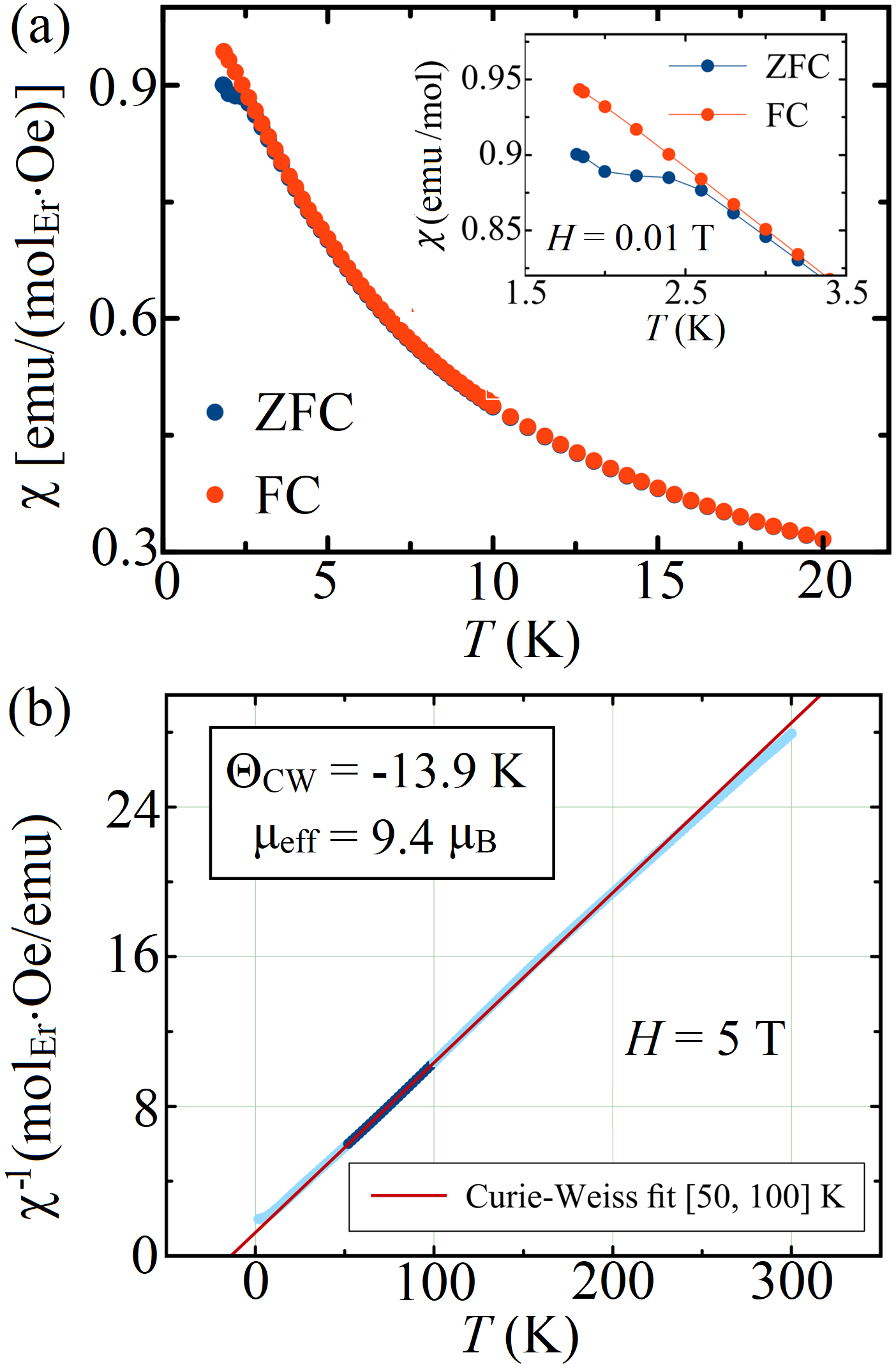}
\caption{(a)~The low-temperature magnetic susceptibility measured from our approximately phase-pure ErMgGaO$_4$ powder sample with a field strength of $H = 0.01$~T, for both FC (red) and ZFC (blue) protocol. The inset shows a close-up of the FC-ZFC bifurcation at $T_g = 2.5$~K. (b)~The inverse magnetic susceptibility measured from our approximately phase-pure ErMgGaO$_4$ powder sample for a field strength of $H = 5$~T. The line shows the Curie-Weiss fit to the inverse susceptibility curve between $T = 50$~and~100~K, yielding $\Theta_{\mathrm{CW}} \approx -14$~K and $\mu_{\mathrm{eff}} \approx 9.4$~$\mathrm{\mu_B}$.}
   \label{Figure2} 
\end{figure}  

\section{Materials Preparation and Characterization}

\subsection{Materials Preparation}

Approximately phase-pure ErMgGaO$_4$ powder was synthesized using a combination of solid-state synthesis and floating-zone image furnace methods. Stoichiometric amounts of Er$_2$O$_3$, MgO, and Ga$_2$O$_3$ were ground together and formed into a rod 10~cm in length and 0.5~cm in diameter. The rod was compressed and subsequently sintered at 1200~$^\circ$C for 48 hours in air. The floating zone growth was performed under 7~atm of oxygen, using a speed of 0.5~mm/h for both the feed and seed rods in an NEC floating zone furnace. The resulting polycrystalline rod was cut into several pieces which were further investigated via both Laue x-ray diffraction and powder x-ray diffraction. While the resulting single crystals were too small to use in neutron scattering experiments, nearly phase-pure polycrystalline regions could be identified and separated, while other regions of the as-grown material tended to be contaminated with an Er$_3$Ga$_5$O$_{12}$ garnet phase around the 1\% level. Several such growths were carried out resulting in a 2.1~gram sample of approximately (97\%) phase-pure powder ErMgGaO$_4$ being extracted, with no detectable Er$_3$Ga$_5$O$_{12}$ impurity. Powder x-ray diffraction of the final extracted sample is shown in Fig.~\ref{Figure1}~(b), with the main contaminant being identified as non-magnetic MgO at the 1-2\% level.

\subsection{Magnetic Susceptibility}

Fig.~\ref{Figure2}~(a) shows the magnetic susceptibility measured from our powder sample of ErMgGaO$_4$ using an MPMS XL SQUID magnetometer with a field strength of $H = 0.01$~T, down to a temperature of $T=1.8$~K. A bifurcation between the FC and ZFC susceptibility is identified as a spin glass transition near $T_g = 2.5$~K, see inset of Fig. ~\ref{Figure2}~(a). This feature was not reported in previous studies on small single crystal samples~\cite{Cai2020}, which may reflect differences in phase purity with the powder samples reported here.  


Fig.~\ref{Figure2}~(b) shows the temperature dependence of the inverse magnetic susceptibility measured from powder ErMgGaO$_4$ with a field strength of $H = 5$~T.  The inverse susceptibility is close to linear, with the small deviation from linearity resulting in a minor temperature dependence for the Curie-Weiss constant extracted from a linear fit. Fitting to a Curie-Weiss relationship from 50 to 100~K gives an antiferromagnetic Curie-Weiss temperature, $\Theta_{\mathrm{CW}} = -14$~K, and an effective paramagnetic moment of $\mu_{\mathrm{eff}} = 9.4$~$\mathrm{\mu_B}$. This compares with Curie-Weiss analysis on earlier ErMgGaO$_4$ samples which yielded estimates for $\Theta_{\mathrm{CW}}$ and $\mu_{\mathrm{eff}}$ of $-30.4$~K and 9.2~$\mathrm{\mu_B}$~\cite{Cava2018}, and the separate estimates of $-33$~K and 10.3~$\mathrm{\mu_B}$~\cite{Cai2020}, for Curie-Weiss fits between 150~K and 300~K. 

\section{High Energy Neutron Spectroscopy and Crystal Electric Field Analysis}

\begin{figure*}[]
\centering
  \includegraphics[width = 0.8\textwidth]{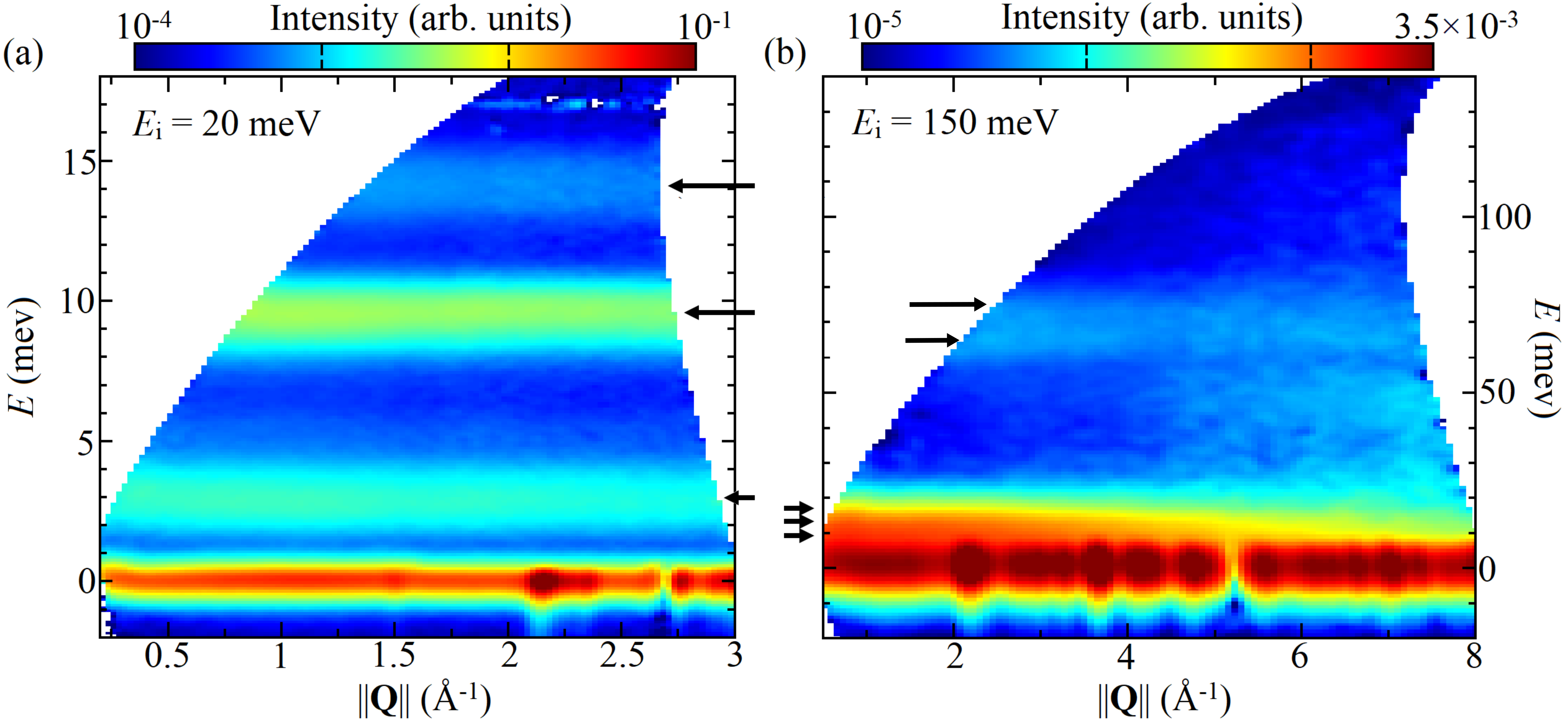}
\caption{Powder-averaged inelastic neutron scattering spectra of ErMgGaO$_4$ at $T = 5$~K for incident neutrons of energy (a)~$E_{\mathrm{i}} = 20$~meV and (b)~$E_{\mathrm{i}} = 150$~meV, with a dataset measured from the empty sample can at $T = 5$~K subtracted for each incident energy. The CEF transitions used to extract the CEF parameters are highlighted by horizontal arrows.}
   \label{Figure3} 
\end{figure*}
\begin{figure*}[]
\centering
  \includegraphics[width = 0.95\textwidth]{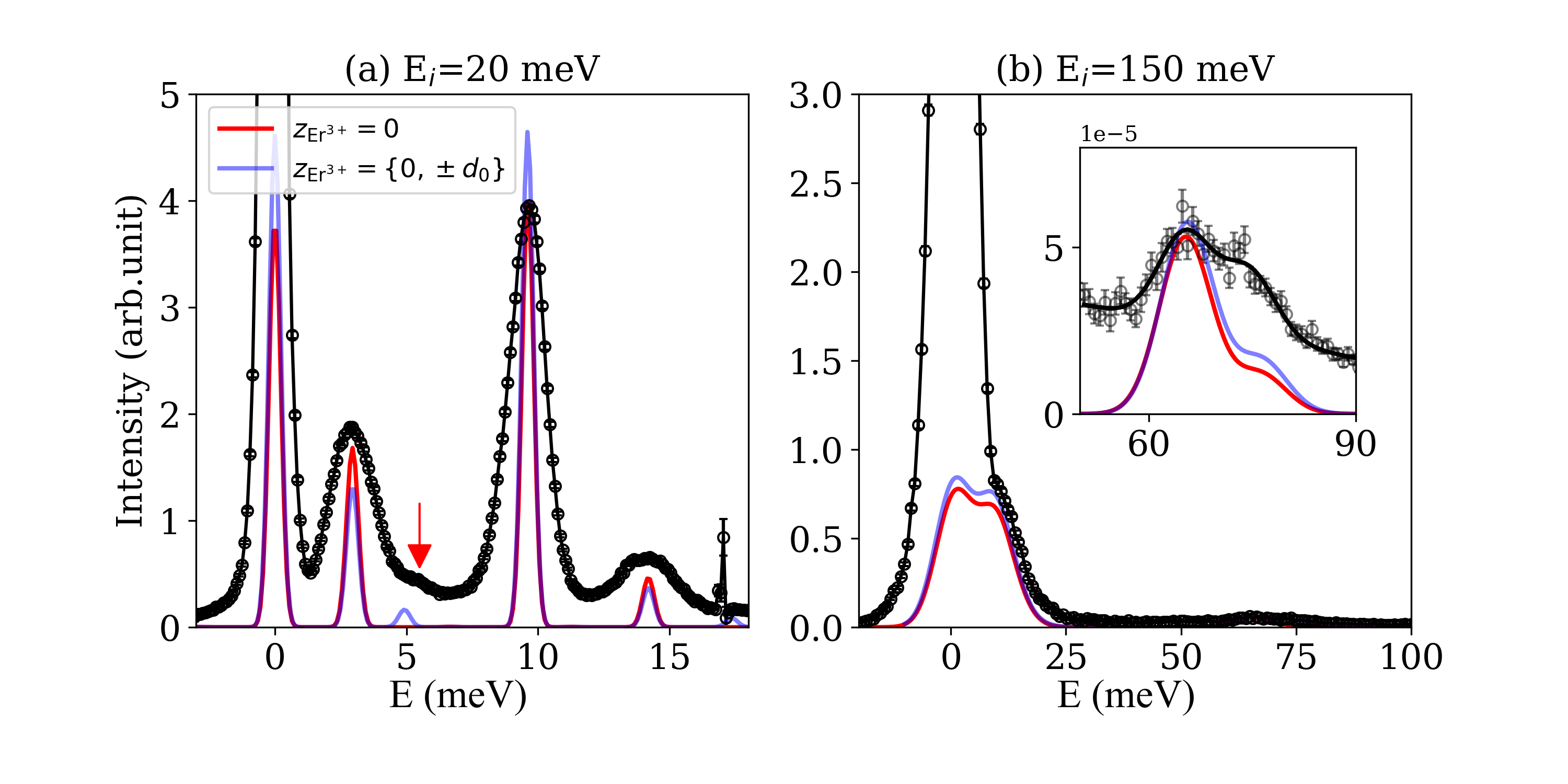}
\caption{Constant $|Q|$ cuts of the (a) $\mathrm{E_i}=20~$meV and (b) $\mathrm{E_i}=150~$meV data. A $|Q|$ integration range of $|Q|=[1.75, 2.5]\AA^{-1}$, and $|Q|=[3.6, 3.85]\AA^{-1}$ are used in (a) and (b), respectively. The red solid lines are calculated CEF excitation spectra convolved with the instrumental resolution at each $\mathrm{E_i}$ by assuming all Er$^{3+}$ to occupy the center of the octahedra. The blue line is obtained by assuming a random occupancy of the centered ($z_{\mathrm{Er}^{3+}}$) and off-centered position ($z_{\mathrm{Er}^{3+}}=\pm d_0\approx\pm 0.1\AA$) with a probability of $0.7$ and $0.3$, respectively (see Appendix for further details). The $\mathrm{E_i}=150~$meV data is enlarged in the inset of (b) to highlight the CEF peak at 70~meV. The black solid line in the inset is fit by the sum of two Gaussian peaks on a linear background. Widths of the two peaks are fixed to the instrumental resolution of $\Delta=9~$meV at $\mathrm{E_i}=150~$meV. }
   \label{Figure4} 
\end{figure*}

  
Microscopically, the crystal electric field (CEF) transitions within Er$^{3+}$ ions, and hence the nature of their single-ion ground state (GS) wave-functions, were studied using thermal time-of-flight (TOF) neutron spectroscopy at the SEQUOIA spectrometer\citep{Granroth2010} as shown in Fig.~\ref{Figure3}. Two incident neutron energies of $\mathrm{E_i}=20~$meV and $\mathrm{E_i}=150~$meV, offering an energy resolution of $\Delta\mathrm{E}\sim$0.5~meV and $\Delta\mathrm{E}\sim$9~meV, respectively (defined as the FWHM of the elastic peaks), are used to access CEF transitions over a wide range of energy transfers, $1~\mathrm{meV}\lesssim \mathrm{E}\lesssim 100~\mathrm{meV}$. Three flat bands of magnetic excitations are clearly observed in Fig.~\ref{Figure3}~(a) at 2.95~meV, 9.6~meV and 14.2~meV, in the low $\mathrm{E_i}=20~$meV data. These excitations, including the one at $\mathrm{E}=$2.95~meV are assigned to be CEF transitions from the GS doublet. The collective magnetic fluctuations {\it within} the GS doublet are found to occur at much lower energy transfers of $\mathrm{E}<1~$meV using cold neutrons (see next section). As shown more clearly by the constant $|Q|$ cuts in Fig.~\ref{Figure4}, the CEF transitions observed in Fig.~\ref{Figure3}~(a), with FWHM's of $\sim2~$meV, are noticeably broader than the instrumental resolution of $\Delta \mathrm{E}\sim0.5~$meV, most likely due to Mg$^{2+}$/Ga$^{3+}$ site-mixing which gives rise to a distribution of local CEF environments. Similar disorder-induced broadening of the CEF transitions has been observed in isostructural YbMgGaO$_4$\citep{Li2017}. Using a higher $\mathrm{E_i}=150$~meV [Fig.~\ref{Figure3}(b)], another CEF peak is observed at $\mathrm{E} \sim70~$meV, with a $|Q|$-dependence clearly different from that of the nearby phonon modes whose intensity is suppressed at small $|Q|$. On the other hand, as shown by the constant $|Q|$ cut of the $\mathrm{E_i}=150~$meV data [Fig.~\ref{Figure4}~(b)], the three lowest CEF transitions in Fig.~\ref{Figure3}~(a) can no longer be resolved due to the coarse energy resolution ($\Delta \mathrm{E}=9$~meV) of the $\mathrm{E_i}=150~$meV data, and appear as a broad shoulder next to the elastic peak. Focusing on the high energy peak at $\mathrm{E}\sim 70~$meV, the inset of Fig.~\ref{Figure4}~(b) shows that it has a width of $\sim$20~meV significantly larger than $\Delta \mathrm{E}=9$~meV. Although it is tempting to attribute the large width to the same disorder-induced spectral broadening as discussed above, such explanation can be ruled out by observing that this effect only contributes to a $\sim$2~meV width for the other three CEF transitions, an order of magnitude smaller than the $20~$meV width observed here. In fact, assuming the same intrinsic width of $2~$meV, a single CEF transition should appear resolution-limited in the $\mathrm{E}_i=$150~meV data with a coarse energy resolution of $\Delta \mathrm{E}=9$~meV. The large width of the peak at $\mathrm{E}\sim$70~meV is therefore more naturally explained by two nearby, overlapping CEF transitions. This is illustrated in the inset of Fig.~\ref{Figure4}~(b), where we fit the data between $50~\mathrm{meV}<\mathrm{E}<90~\mathrm{meV}$ to a sum of two Gaussians and a linear background (black solid line), centered at 65(2)~meV and 75(2)~meV, whose widths are constrained to be the instrumental resolution of $\Delta \mathrm{E}=9~$meV. The experimentally determined energies, and peak intensities of the observed CEF transitions (relative to the intensity of the CEF transition at $\mathrm{E}=$9.6~meV), are summarized in Table~\ref{TableI}.

\begin{table}[]
\begin{tabular}{|c|c|c|c|c|}
\hline
 Transition & $\mathrm{E}_i^\mathrm{exp}~(\mathrm{meV}$) & $I_i^\mathrm{exp}$ & $\mathrm{E}_i^\mathrm{theo}~(\mathrm{meV}$) & $I_i^\mathrm{theo}$ \\ \hline

\begin{tabular}[c]{ccccc}  \end{tabular}          
1& $2.95(7)$ & $0.44(5)$ & 2.95 & 0.42\\ \hline

\begin{tabular}[c]{ccccc}  \end{tabular}          
2& $9.6(1)$ & $1$ & 9.62 &1 \\ \hline

\begin{tabular}[c]{ccccc}  \end{tabular}          
3& $14.2(2)$ & $0.12(2)$ &14.20 &0.12\\ \hline

\begin{tabular}[c]{ccccc}  \end{tabular}          
4& $65(2)$ & $0.04(2)$ &65.24 &0.09\\ \hline

\begin{tabular}[c]{ccccc}  \end{tabular}          
5& $75(2)$ & $0.032(15)$&76.06 & 0.02 \\ \hline
\end{tabular}
\caption{Experimentally determined energies, $\mathrm{E}_i^\mathrm{exp}$, and intensities of the CEF transitions, $I_i^\mathrm{exp}$ (intensity of the second transition, $I_2$, has been normalized to 1). $\mathrm{E}_i^\mathrm{exp}$ and $I_i^\mathrm{exp}$ are obtained by fitting to the constant $|Q|$ cuts in Fig.~\ref{Figure4} as described in the Appendix. The corresponding theoretical values, $\mathrm{E}_i^\mathrm{theo}$ and $I_i^\mathrm{theo}$, determined by Eq.~\eqref{eq:1} are given in the last two columns.}
\label{TableI}
\end{table}

\begin{figure*}[]
\centering
  \includegraphics[width = 0.95\textwidth]{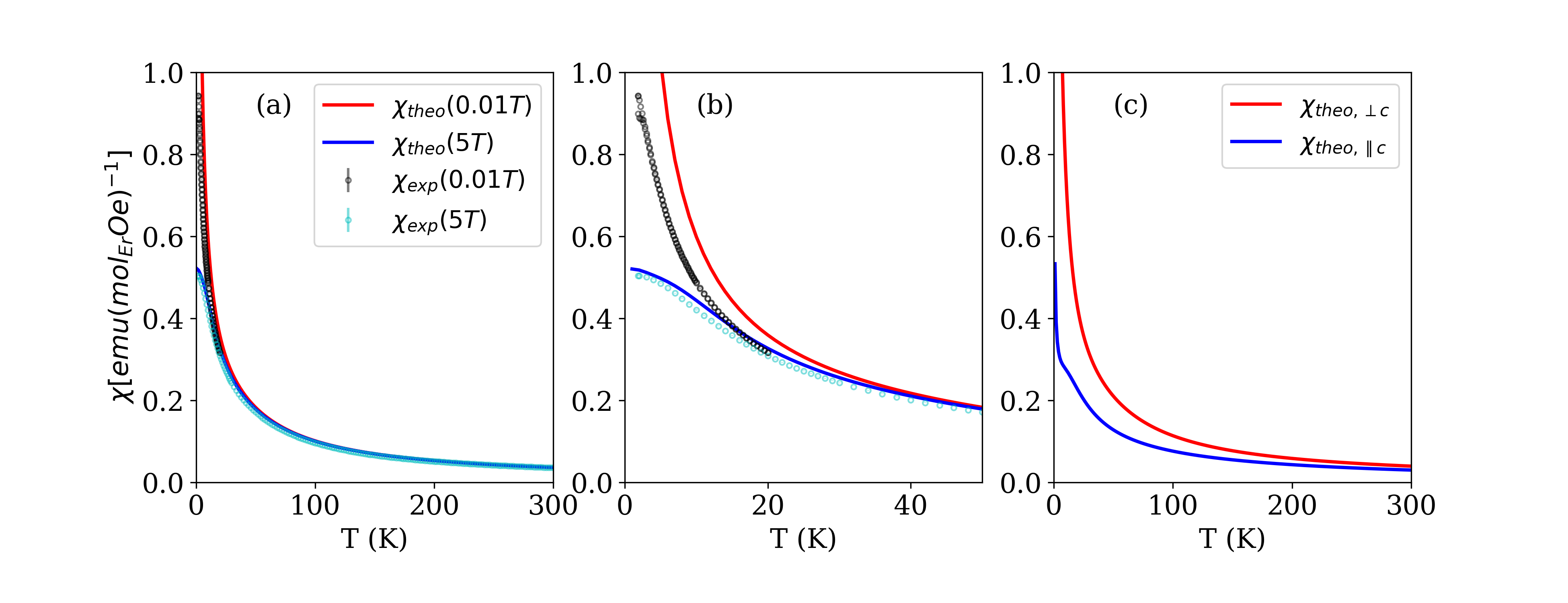}
\caption{(a,b) Temperature-dependent magnetic susceptibilities, $\chi_\mathrm{exp}=\frac{M}{H}$, of a powder sample of ErMgGaO$_4$ measured with an external field of $H=0.01~$T (black circle) and $H=5~$T (cyan circle). Theoretically calculated magnetic susceptibilities at the same fields, $\chi_\mathrm{theo}$, are shown in solid lines. (b) is the same as (a) but zooms into the region with $\mathrm{T<50}~$K. (c) Theoretically predicted in-plane ($\chi_{\perp c}$) and out-of-plane ($\chi_{\parallel c}$) magnetic susceptibilities. }
   \label{Figure5} 
\end{figure*}  

The CEF levels within the $J=\frac{15}{2}$ GS multiplet of the  Er$^{3+}$ ions determined by Hund's rules are modelled using the Stevens operator formalism with the following CEF Hamiltonian:
\begin{equation} \label{eq:1}
    \mathcal{H}_\mathrm{CEF}^{D_{3d}} = B_2^0 \hat{O}_2^0 + B_4^0 \hat{O}_4^0 + B_4^3 \hat{O}_4^3 + B_6^0 \hat{O}_6^0 + B_6^3 \hat{O}_6^3 + B_6^6 \hat{O}_6^6 ,\; 
\end{equation} 
where $\hat{O}_m^n$ is the $m^{th}$-order Stevens operator written in terms of the total angular momentum operators, with the corresponding CEF parameter, $B_m^n$. Other $B_m^n$ not included in Eq.~\eqref{eq:1} are zero for the $D_{3d}$ point group appropriate for ErMgGaO$_4$. Fitting to the experimentally determined energies and intensities of the CEF transitions is carried out using a generalized simulated annealing algorithm with the following cost function:
\begin{equation} \label{cost}
     \mathcal{C}= \log_{10}\left(1+\frac{1}{N}\sum_{i=1}^N \left|\frac{X_i^\mathrm{exp}-X_i^\mathrm{theo}}{\sigma_i}\right|\right) ,\; 
\end{equation}
with experimental observables $\{X_i^\mathrm{exp}\}=\{\mathrm{E}_1^{exp}, \mathrm{E}_2^{exp}, \mathrm{E}_3^{exp}, \mathrm{E}_4^{exp}, \mathrm{E}_5^{exp}, I_1^{exp}, I_3^{exp},I_4^{exp},I_5^{exp}\}$ from Table~\ref{TableI}, and their corresponding uncertainties $\{\sigma_i\}=\{0.1, 0.1, 0.2, 2, 2, 0.05, 0.02,0.02,0.02\}$. Since the difference between theory and experiment, $|X_i^\mathrm{exp}-X_i^\mathrm{theo}|$, depends on the fitting parameters, $B_m^n$, in a highly non-linear way, and can vary by orders of magnitude across the entire parameter space, $\log_{10}$ is added to smooth the $\mathcal{C}$ landscape and ensure the convergence of the annealing process. 

Two sets of $\{B_m^n\}$'s obtained from annealing process show excellent agreement with the experimental data (See Appendix for further details). Simulated INS spectrum using the better set, $\{B_m^n\}=\{5.53\times10^{-1}, 2.67\times 10^{-3}, 3.83\times 10^{-2}, 3.27\times10^{-5}, -6.59\times10^{-5}, 7.82\times10^{-5}\}$~meV (in the order shown in Eq.~\eqref{eq:1}) are shown in Fig.~\ref{Figure4} as red solid lines after convolving with the instrumental resolution, which capture the experimentally measured spectra reasonably well. The temperature-dependent powder-averaged magnetic susceptibilities at 0.01~T and 5~T calculated using the same set of $\{B_m^n\}$, $\chi_\mathrm{theo}$ [See Fig.~\ref{Figure5}~(a),(b)], also compare favorably with the experimental data ($\chi_\mathrm{exp}$) at both fields, showing excellent agreement at $\mathrm{T}\gtrsim 30~$K with slight deviation at low temperatures. In Fig.~\ref{Figure5}~(c), we simulate the single crystal magnetic susceptibilities revealing an XY-like single-ion anisotropy at all temperatures. The temperature-dependent single-crystal susceptibility can be used to distinguish between different sets of CEF parameters that otherwise predict similar low-temperature INS spectra and powder-averaged susceptibilities (see Appendix).

The GS doublet wave-functions predicted by our model are:
\begin{align} \label{wavefunction}
    |\pm\rangle_0=& 0.188\left|\mp\tfrac{13}{2}\right\rangle
    \pm 0.353\left|\mp\tfrac{7}{2}\right\rangle \nonumber\\
    & +0.666 \left|\mp\tfrac{1}{2}\right\rangle \mp 0.433 \left|\pm\tfrac{5}{2}\right\rangle +0.457\left|\pm\tfrac{11}{2}\right\rangle \; 
\end{align},
with the in-plane and out-of-plane components of the $g-$tensor given by $g_{xy}=g_J\left\langle +\left|\hat{J}_+\right |-\right\rangle=8.07$ and $g_{z}=2g_J\left\langle +\left|\hat{J}_z\right |+\right\rangle=1.75$, respectively, implying an easy-plane type anisotropy at low temperatures [see Fig.~\ref{Figure5}~(c)]. $g_{xy}\gg g_z$ predicted here also implies a strongly XY-like pseudo-spin interaction by projecting a bilinear spin interactions ($|\Delta m_J|=1$) onto the GS doublet, although the precise form of the pseudo spin Hamiltonian (see next section) is also influenced by multipolar-interactions with larger $|\Delta m_J|<7$ for $f-$electron systems\citep{RevModPhysmultipolar, Yb2Pt2Pbscience}, and the presence of a low-lying CEF level \citep{VirtualCEF} at $\mathrm{E}=2.95~$meV with the following wave-functions:
\begin{align} \label{wavefunction2}
|\pm\rangle_1
&= \;\pm0.077\left|\mp\tfrac{13}{2}\right\rangle
   +0.195\left|\mp\tfrac{7}{2}\right\rangle
   \pm0.551\left|\mp\tfrac{1}{2}\right\rangle \nonumber\\
&\quad
   +0.220\left|\pm\tfrac{5}{2}\right\rangle
   \mp0.777\left|\pm\tfrac{11}{2}\right\rangle \; .
\end{align}. This is to be contrasted with the result of a point-charge model predicting $|\pm\rangle _\mathrm{0, pc}\approx \left|\pm\frac{15}{2}\right\rangle$ , with a $|\Delta m_J|$ larger than the maximum allowed $|\Delta m_J|=7$ in $f-$electron systems \citep{Cai2020}, and therefore a strictly Ising-like pseudo-spin interaction precluding any collective excitations within the GS doublet. (Any transverse pseudo-spin interaction within a point charge model can only come from virtual excitation to the excited CEF levels, which are predicted to occur at much higher energies)

\begin{figure}[]
\centering
  \includegraphics[width = 0.5\textwidth]{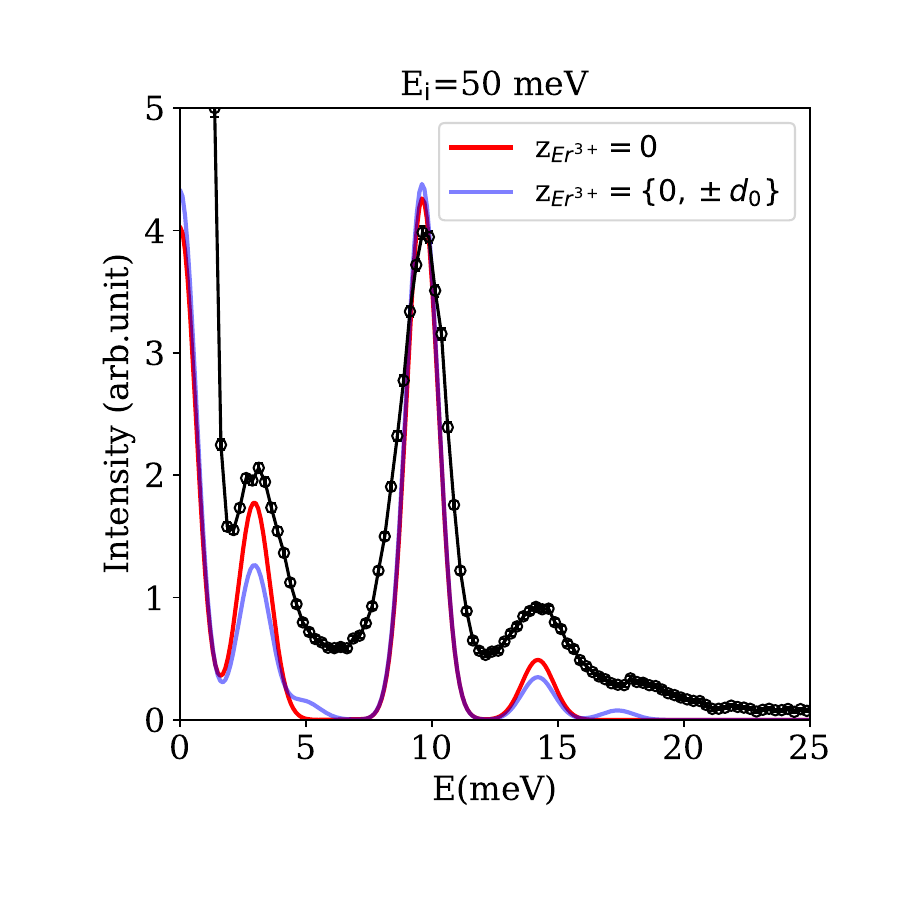}
\caption{Constant $|Q|$ cuts of the $\mathrm{E_i}=50~$meV data using a $|Q|$ integration range of $|Q|=[3, 4]\AA^{-1}$. The red and blue solid lines are the same calculated CEF excitation spectra as Fig.~\ref{Figure4}, obtained by assuming a single Er$^{3+}$ position, and a random site occupancy at $z=0$ and $z=\pm d_0\approx0.1\AA$, respectively. See Appendix for further details of the CEF calculations. }
   \label{Ei50meV}
\end{figure}

Despite the satisfactory agreements between theory and experiment shown in Fig.~\ref{Figure4} and Fig.~\ref{Figure5}, we note two salient features in the data not captured by our model using a single set of CEF parameters. First, the simulated $\mathrm{E_i}=20~$meV spectrum is clearly sharper than the data [Fig.~\ref{Figure4}~(a)]. In addition, scrutinizing the data in Fig.~\ref{Figure4}~(a) also reveals a weak bump at $\sim$5~meV (indicated by a red vertical arrow) absent in our model. Any attempts to reproduce this feature significantly reduces the quality of the fit. As discussed above, both caveats might be related to the presence of local structural disorder, and the $B_m^n$'s obtained above (by fitting the INS spectra to a single set of CEF parameters) reflects only the average CEF environment. In fact, as shown by the blue solid lines in Fig.~\ref{Figure4} and discussed in more details in the Appendix, a model incorporating the \textit{experimentally} observed structural disorder, namely a random Er$^{3+}$ off-centering displacement \citep{Cava2018}, could account for the observed weak peak at $\sim 5~$meV without affecting the high energy CEF spectrum [Fig.~\ref{Figure4}~(b)] or the magnetic susceptibility. Notably, as shown in Fig.~\ref{Figure4}~(a), the same CEF model with disorder also predicts a weak peak at $\mathrm{E}\sim17~\mathrm{meV}$, absent in the CEF spectrum with a single Er$^{3+}$ position. Presence of the $\mathrm{E}\sim17~\mathrm{meV}$ excitation, unobservable in the $\mathrm{E_i}=20~$meV data due to kinematic constraint, is confirmed by using a larger $\mathrm{E_i}=50~$meV (Fig.~\ref{Ei50meV}) as a shoulder next to the main $\mathrm{E}=14.2~$meV peak, highlighting the influence of structural disorder on the CEF spectra in ErMgGaO$_4$.

\section{Low Energy Neutron Scattering Measurements}

\begin{figure}[h!]
  \centering
  \includegraphics[width=0.5 \textwidth]{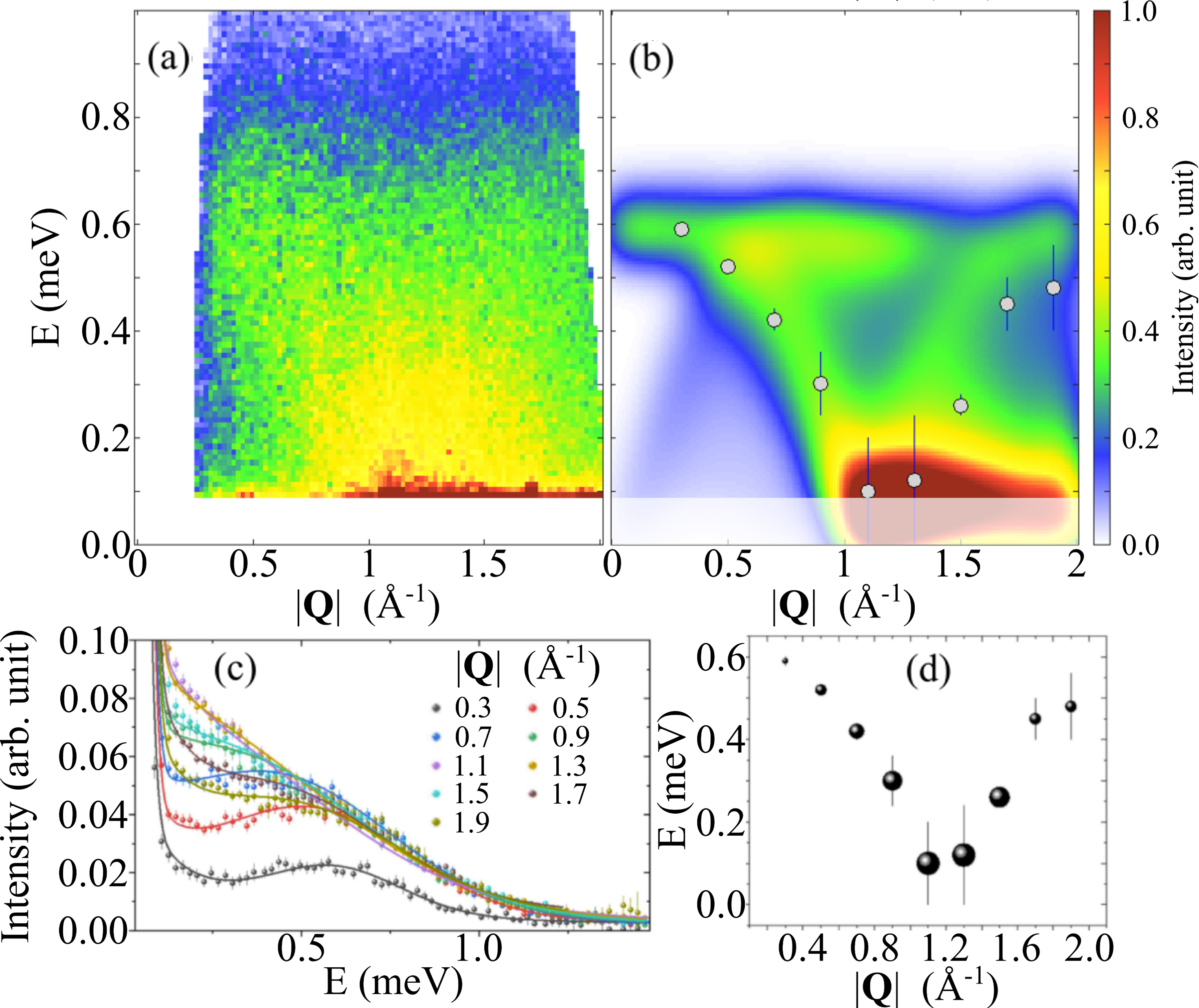}
  \caption {(a)~The full inelastic scattering data set for powder ErMgGaO$_4$ at T=0.125 K is shown. (b)~The best fit LSWT description, via SpinW, of the full inelastic neutron scattering data set, as calculated with the $J_1$-$J_2$ - $\Delta$ Hamiltonian and using $\frac{J_2}{J_1}=0.13$ and $\Delta=0.4$ is shown. (c) Shows constant $|$Q$|$ cuts through the data, fit to Eq.~\eqref{SQw}.  The resulting dispersion of $E_c$ (from fits using Eq.~\eqref{SQw}) vs $|$Q$|$ is shown in (d), where the size of the data points is proportional to the amplitude A$_3$ of the inelastic component of the scattering in Eq.~\eqref{SQw}.  This dispersion is overlaid on the best fit LSWT calculation in (b), which  then indicates the extent to which the LSWT calculation captures the inelastic spectral weight.}

   \label{lowEi} 
\end{figure}

Low energy time-of-flight neutron scattering measurements were performed on a 2~gram sample of approximately phase pure ErMgGaO$_4$, using the IN6-Sharp time-of-flight spectrometer at the Institut Laue-Langevin in France~\cite{ILL_experiment_DOI}. This set of measurements employed an incident neutron energy of 3.12~meV.  The powder sample was loaded in a He atmosphere in a dilution refrigerator to access temperatures well below 1~K, with a base temperature of $\sim$ 0.125~K.  This set of neutron measurements, shown in Fig.~\ref{lowEi}(a), is complementary to the previously described SEQUOIA measurements by covering the low energy range from $-1$~meV~$ < \Delta E < 2$~meV. 

The measured dynamic magnetic spectral weight for ErMgGaO$_4$ at the base temperature for the IN6-Sharp experiment, $T=0.125$~K, is shown in Fig.~\ref{lowEi}(a) and (c).  It can be described in terms of a broad continuum of scattering with a bandwidth of $\sim$ 0.8~meV, and is mostly contained within the $|$Q$|$ range from $0.5 ~\AA^{-1} < |Q| < 2 ~\AA^{-1}$.  The overall bandwidth of the magnetic excitation spectrum is an estimate of the overall interaction strengths in magnetic materials, similar to $\Theta_{\mathrm{CW}}$. As discussed, our low energy neutron spectroscopy shows that the top of the spin excitation band is $\sim$ 0.8~meV ($\sim$ 10~K), and is therefore reasonably close to the $\Theta_{\mathrm{CW}}$ estimate made from our susceptibility measurements. We thus conclude that an estimate of $\Theta_{\mathrm{CW}}$, to be in the range from $-10$~K to $-14$~K, is both reasonable and useful.

\begin{figure}[]
  \centering
  \includegraphics[width=0.5 \textwidth]{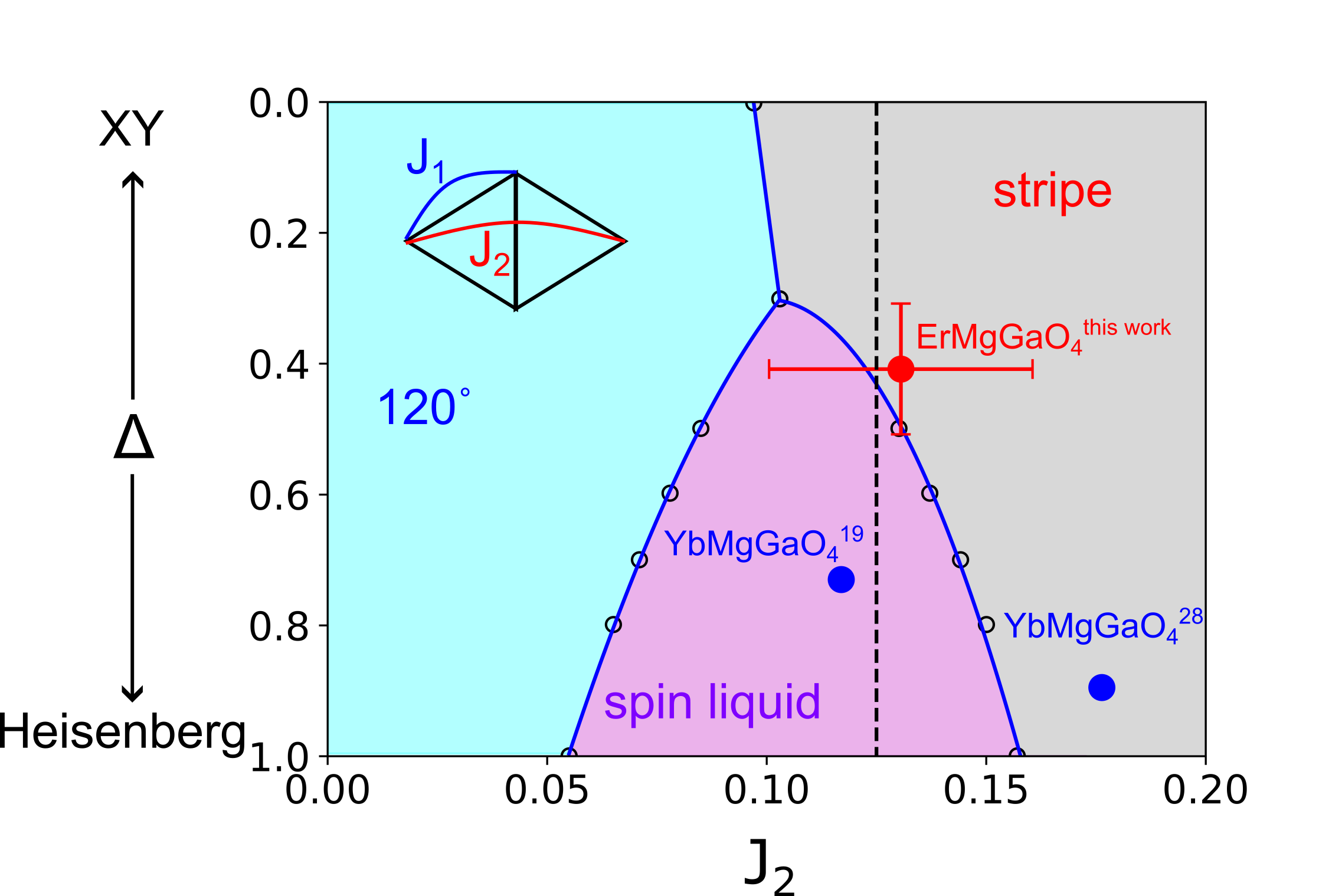}
  \caption {The theoretical $\frac{J_2}{J_1}$ - $\Delta$ ground state phase diagram for the effective spin-1/2 Hamiltonian [Eq.~\eqref{eq:2}] on a triangular lattice, relevant to ErMgGaO$_4$ and YbMgGaO$_4$, with $J_{z\pm}$ = $J_{\pm\pm}$ = 0~\cite{Chernyshev2017} is shown. $J_1$ and $J_2$ are the nearest neighbour and next near neighbour exchange interactions within the triangular plane, as indicated in the inset to the figure, and $J_1$=1.  The red and blue points show the locations of the estimates for ErMgGaO$_4$ and YbMgGaO$_4$ in this ground state phase diagram according to the analyses in both this work and in Ref.~\cite{Steinhardt2021, Mourigal2018}, respectively.}
   \label{phase diagram} 
\end{figure} 

\subsection{Comparison to Linear Spin Wave Theory}

This measured inelastic scattering from powder ErMgGaO$_4$ at T=0.125 K is shown in Fig.~\ref{lowEi}(a) and (c), where the full data set is shown in (a) while cuts through this data at varying $|$Q$|$ are shown in (c).  The $|$Q$|$ cuts as a function of energy in Fig.~\ref{lowEi}(c) were then fit to the phenomenological form:

\begin{align}\label{SQw}	
S(Q, E) & = A_1(Q) \exp(-(\frac{E}{\sqrt{2}w_1})^2)  \nonumber\\ & + \bigg\{\frac{2A_2}{\pi} [\frac{Ew_2^2}{4E^2+w_2^2}]+A_3\exp(-[\frac{E-E_c}{w_3}]^2)\bigg\}  \nonumber\\ &
\times \frac{1}{1-e^{-E/k_{\mathrm{B}}T}} ~~.
\end{align}

for the purposes of broadly describing this spectral weight.

The first two terms in Eq.~\eqref{SQw} model the elastic and quasi-elastic components of the scattering, respectively.  The elastic component has an amplitude $A_1(Q)$, and a width $w_1$ given by the energy resolution. The quasi-elastic component has an amplitude ${\frac{2A_2}{\pi}}$ and its width is governed by $w_2$.  The third Gaussian term, centred at $E_C$, gives the inelastic peak, with an amplitude $A_3$ and a width governed by $w_3$.  Both the quasi-elastic and inelastic terms are modulated by an appropriate Bose factor.  The overall description of the $|$Q$|$ cuts in Fig.~\ref{lowEi} (c) by the fits using Eq.~\eqref{SQw} are very good. The $E_C$ vs $|$Q$|$ dispersion resulting from these fits is shown in Fig.~\ref{lowEi}~(d), where the size of the data points is proportional to the amplitude of this inelastic contribution to the scattering, $A_3$.  This shows that the experimental continuum of inelastic scattering is largely contained within the fan of points describing the $E_C$ vs $|$Q$|$ dispersion.

The following anisotropic exchange spin Hamiltonian on a two-dimensional triangular lattice has been used extensively in LSWT calculations to model low lying spin excitations in the well studied Yb$^{3+}$-based triangular lattice materials, including YbMgGaO$_4$ and YbZnGaO$_4$~\cite{Steinhardt2021}:

\begin{align}\label{eq:2}	
H & = \sum_{\langle ij \rangle_{NN}} \bigg\{ J_1(S^{x}_iS^{x}_j + S^{y}_iS^{y}_j + \Delta S^{z}_iS^{z}_j)  \nonumber \\
& + 2J_{\pm\pm}\big[(S^{x}_iS^{x}_j - S^{y}_iS^{y}_j)\text{cos}\tilde{\phi_{a}} - (S^{x}_iS^{y}_j + S^{y}_iS^{x}_j)\text{sin}\tilde{\phi_{a}}\big]  \nonumber \\
& + J_{z\pm}\big[(S^{y}_iS^{z}_j + S^{z}_iS^{y}_j)\text{cos}\tilde{\phi_{a}} - (S^{x}_iS^{z}_j + S^{x}_iS^{z}_j)\text{sin}\tilde{\phi_{a}}\big] \bigg\}  \nonumber \\
& + \sum_{\langle ij \rangle_{NNN}} J_{2} (S^{x}_iS^{x}_j + S^{y}_iS^{y}_j + \Delta S^{z}_iS^{z}_j)~~.
\end{align}

It includes both nearest-neighbor (NN) and next-nearest-neighbor (NNN) exchange interactions, although for simplicity the $\Delta$ anisotropy parameter is often assumed to be the same for both NN and NNN interactions, and both $J_{\pm\pm}$ and $J_{z\pm}$ are restricted to NN interactions only.

\begin{figure*}[]
\centering
  \includegraphics[width=0.96 \textwidth]{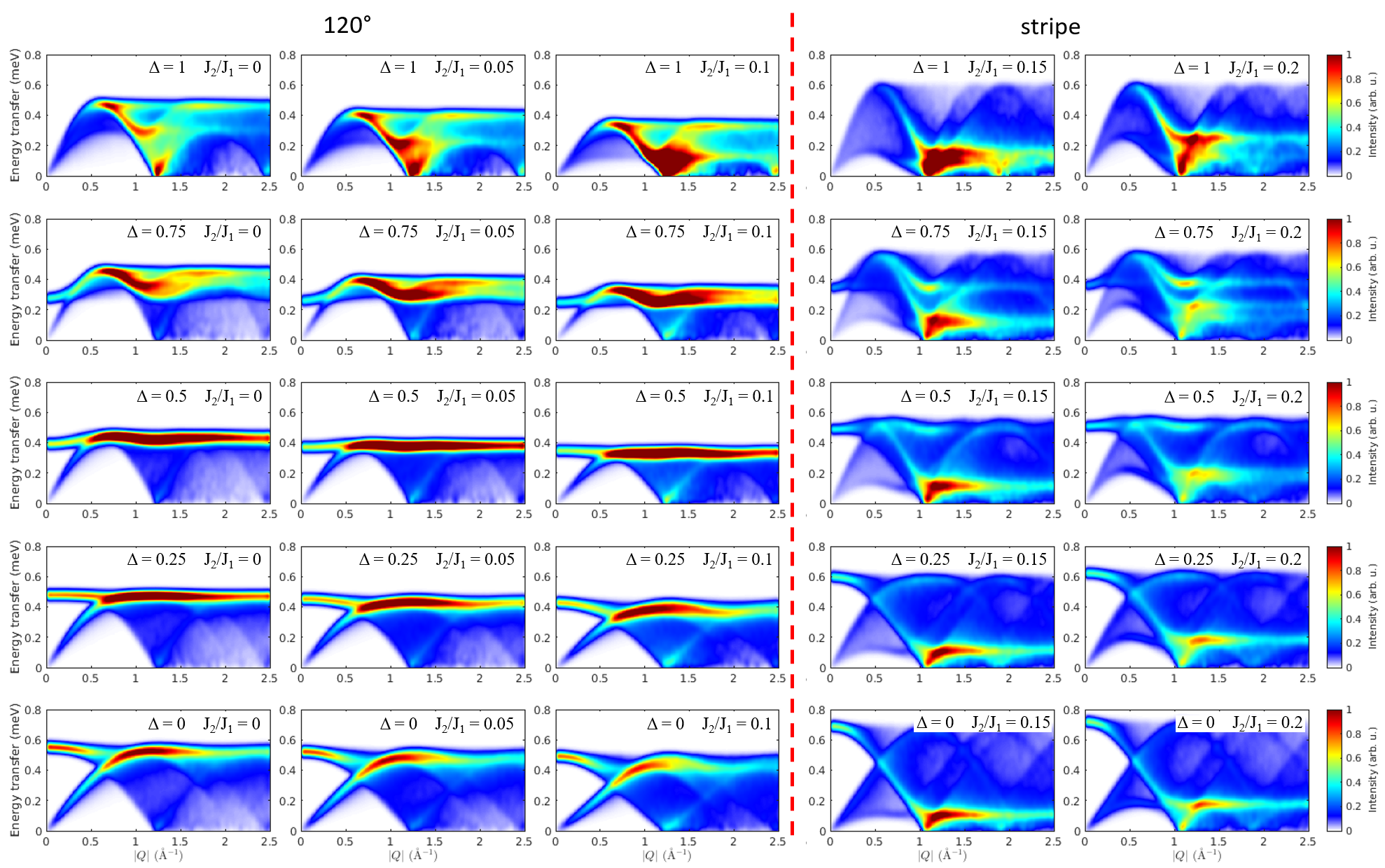}
\caption{The powder-averaged dynamic structure factor calculated via LSWT and the SpinW package, at $T = 0$~K is shown.  These calculations use the Hamiltonian of Eq.~\eqref{eq:2} for $J_2/J_1$ ranging from 0 to 0.2, and $\Delta$ ranging from 0 (XY anisotropy) to 1 (isotropic Heisenberg).  The red dashed line indicates $\frac{J_2}{J_1}$=0.126, where the classical ground state transitions between an ordered non-collinear 120$^\circ$ phase and an ordered collinear stripy phase.}
   \label{spinwaveSQw} 
\end{figure*}  

The theoretical phase diagram relevant to this Hamiltonian has been well-explored and the expectations for the zero temperature phases within the $J_2$/$J_1$ vs. $\Delta$ plane are shown in Fig.~\ref{phase diagram} ~\cite{Chernyshev2025} . This shows the phase behavior expected both classically and quantum mechanically, through density matrix renormalization group (DMRG) techniques.  Here, the bond dependent terms in Eq.~\eqref{eq:2}, $J_{\pm\pm}$ and $J_{z\pm}$, have both been set equal to zero, and, in what follows, we will continue to employ the simplification $J_{\pm\pm} = J_{z\pm}=0$. We will be describing powder inelastic scattering for which the bond-dependent terms will not substantially impact the description of the data, so long as they are relatively weak.

Classically, there is a direct transition between an ordered non-collinear 120$^\circ$ phase and an ordered collinear stripy phase at $J_2$/$J_1$ $\sim$ 0.126, independent of $\Delta$.  This is indicated as the vertical dashed line in Fig. ~\ref{phase diagram}.  However the quantum version as calculated with DMRG introduces an intermediate quantum spin liquid phase for 0.06 $\le$ $J_2$/$J_1$ $\le$ 0.16, between the two ordered N\'eel states.

We have performed LSWT calculations ~\cite{Toth2015} to explore the anticipated dynamic spectral weight for the anisotropic $J_1$-$J_2$ model, as expressed by the Hamiltonian Eq.~\eqref{eq:2}.  A subset of these calculations are shown in Fig.~\ref{spinwaveSQw}, which shows the powder-averaged dynamic structure factors as $|$Q$|$ vs. $E$ for values of $J_2$/$J_1$ between 0 and 0.2, and for $\Delta$ between 0 (extreme XY) and 1 (Heisenberg). One can see that while there is some dependence on the details of the microscopic parameters, the more substantial differences are associated with whether the classical ground state is an ordered non-collinear 120$^\circ$ N\'eel phase (as it is to the left of the red dashed line in Fig.~\ref{spinwaveSQw}) or an ordered collinear stripy phase (as it is to the right of the red dashed line in Fig.~\ref{spinwaveSQw}). Importantly, except for the Heisenberg $\Delta$=1 case, the dynamic spectral weight calculated using LSWT for a 120 degree ground state gives spectral weight concentrated at high energies and is relatively dispersionless, whereas the spectral weight is concentrated in a flat low energy mode for the stripy phase.

In order to approximately place ErMgGaO$_4$ in the theoretical phase diagram in Fig.~\ref{phase diagram}, these LSWT calculations are compared to the measured dynamic structure factor for ErMgGaO$_4$ powder as shown in Fig.~\ref{lowEi}.  The best attempt to model the full data set at T=0.125 K within LSWT is shown in Fig.~\ref{lowEi}~(b), overlaid with the experimentally determined $E_C$ vs $|$Q$|$ dispersion from Fig.~\ref{lowEi}(d). The LSWT calculations qualitatively capture the observed low energy spin excitations, including the bandwidth and the overall shape, as well as an accumulation of low energy scattering at $|$Q$|>$1 A$^{-1}$ and $E\lesssim 0.1~$meV. This exercise allows us to estimate the microscopic spin Hamiltonian parameters for ErMgGaO$_4$ at T=0.125~K, to be $\frac{J_2}{J_1}=0.13 \pm 0.03$ and $\Delta=0.4 \pm0.1$. The approximate placement of ErMgGaO$_4$ in the theoretical phase diagram in Fig.~\ref{phase diagram} has been indicated by a filled red circle together with the errorbars.  For reference, we have also indicated two existing estimates for the position of YbMgGaO$_4$ within this same phase diagram ~\cite{Steinhardt2021, Mourigal2018}. One can see that this places ErMgGaO$_4$ immediately adjacent to the QSL phase of the zero temperature theoretical phase diagram, within the ordered stripy N\'eel phase. In addition to structural disorder, the enhanced quantum fluctuations in close proximity to the QSL phase also provides a natural explanation for the observation of a significantly broader and more diffuse excitation spectrum than that predicted by LSWT.

Finally, we note that the presence of the low energy, first excited state CEF levels $\sim$ 3 meV above the ground state is likely to effect the details of both the exchange and anisotropy terms in the Hamiltonian, Eq.~\eqref{eq:2},  due to virtual CEF transitions.  These effects have been well studied in the pyrochlore Er$_2$Ti$_2$O$_7$~\cite{Savary2012, Rau2016} where a first excited CEF doublet is $\sim$ 6 meV above the ground state, roughly double the CEF energy gap above the ground state for the case of ErMgGaO$_4$.

\subsection{Temperature Dependence of Low Energy Spin Excitations}

We can phenomenologically describe the inelastic spectral weight at all measured $|Q|$ and $E$ using the sum of two damped harmonic oscillators, with each DHO expressed analytically as:  
\begin{align}\label{eq:3}	
S(Q, E) & = A(Q) \left[\frac{w}{(E-E_{c})^2 + w^2} - \frac{w}{(E+E_{c})^2 + w^2}\right]  \nonumber \\
& \times \frac{1}{1-e^{-E/K_{\mathrm{B}}T}} ~~.
\end{align}

\begin{figure}[]
  \centering
  \includegraphics[width=0.45\textwidth]{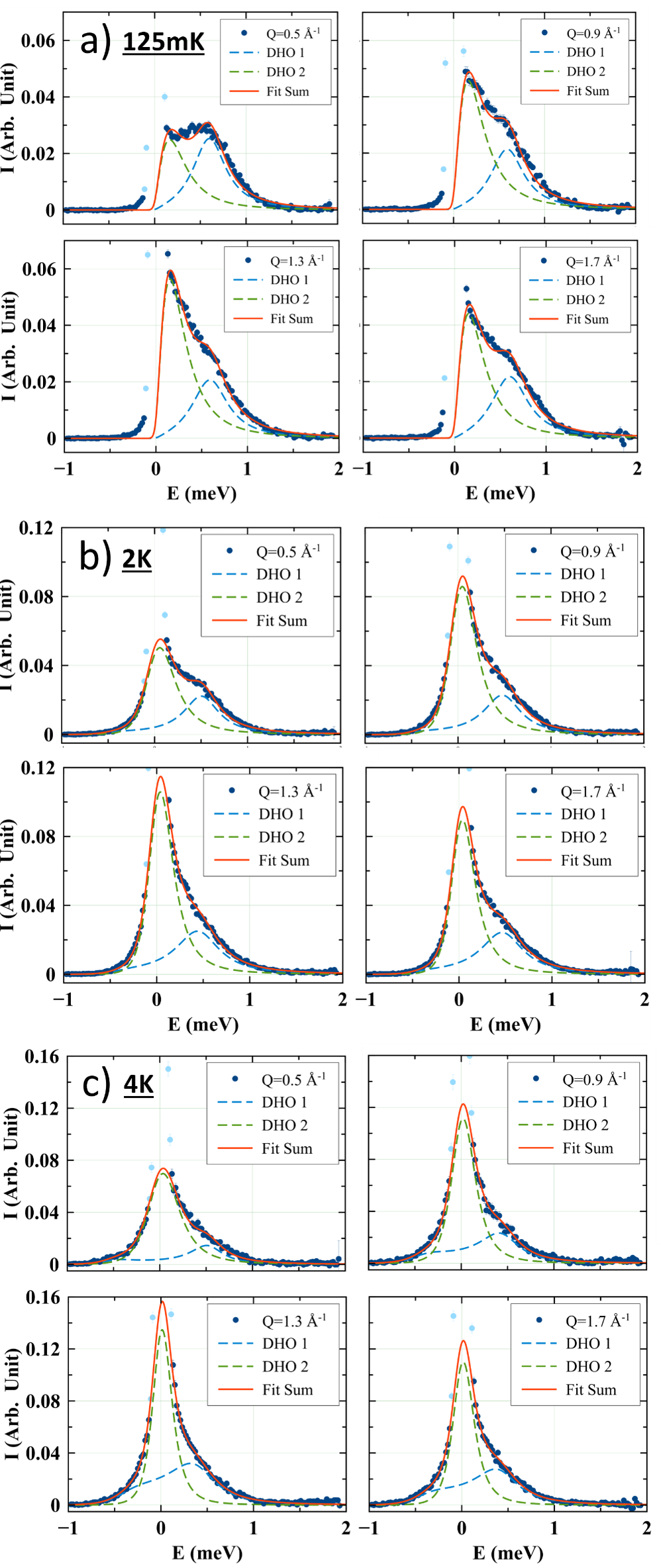}
  \caption{Double damped harmonic oscillator fits to the inelastic neutron scattering spectra measured from powder ErMgGaO$_4$ along $Q = 0.5$, 0.9, 1.3, and 1.7~$\angstrom^{-1}$, plotted for (a)~$T = 0.125$~K, (b)~$T = 2$~K, (c)~$T = 4$~K. Specifically, we show the measured data for an integration in $Q$ centered on the labeled $Q$ value in each case ($Q = 0.5$, 0.9, 1.3, and 1.7~$\angstrom^{-1}$), with a $Q$-width of 0.1~$\angstrom^{-1}$ used for the integration in each case. The energy center of DHO2 was constrained to 0.01~meV. Elastic scattering between $-0.08$~meV and 0.08~meV (light blue data points) was ignored for each fit. The error bars for the data are within the point size.} 
  
   \label{Tempdep_fit} 
\end{figure} 

\begin{figure}[]
  \centering
  \includegraphics[width=0.5 \textwidth]
  {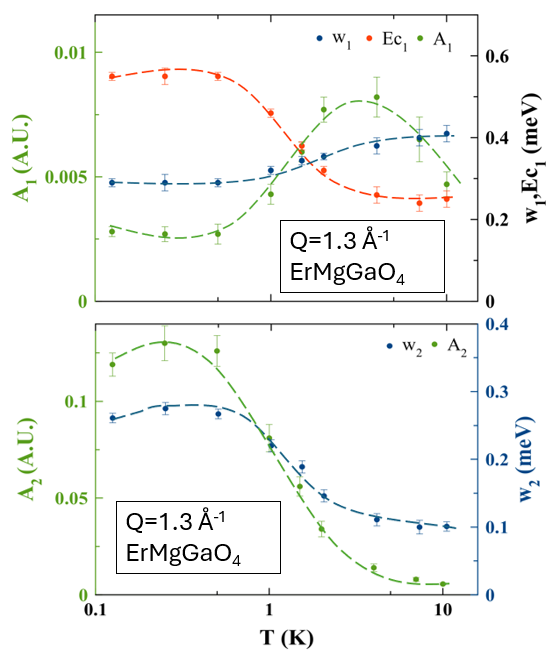}
  \caption{The temperature dependence of the five parameters used in the double-DHO description of the inelastic neutron scattering signal measured from ErMgGaO$_4$ in this work. Specifically, we show (a)~the amplitude, energy-width, and energy-centre of the higher energy contribution (DHO1) and the (b) amplitude and energy-width of the quasi-elastic contribution (DHO2) as averaged over  $Q = [1.1, 1.3]$~$\angstrom^{-1}$ in each case. The dashed lines are guides to the eye.}
   \label{Tempdep_parameters} 
\end{figure}

\begin{figure}[]
  \centering
  \includegraphics[width=0.5 \textwidth]{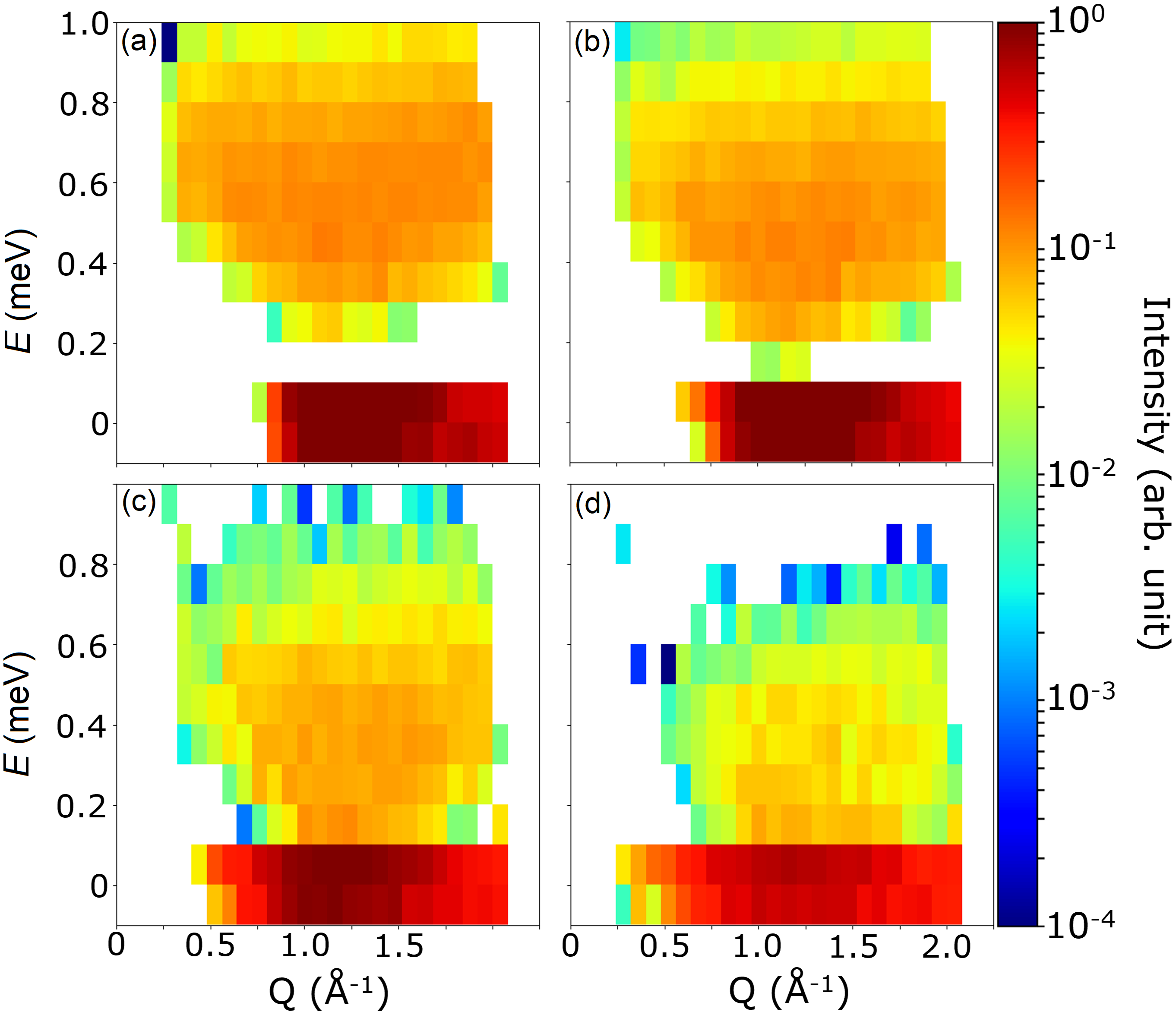}
  \caption{The temperature-subtracted time-of-flight neutron scattering signal measured from ErMgGaO$_4$ in this work. Specifically, we show the measured signal from ErMgGaO$_4$ at (a) $T = 0.125$~K, (b) $T = 1.5$~K, (c) $T = 4$~K, and (d) $T = 10$~K with a $T = 100$~K dataset subtracted in each case, plotted on logarithmic intensity scale with negative net-intensity shown as white.}
   \label{tempsubtraction} 
\end{figure}  	

Each DHO lineshape has its own $Q$-dependent amplitude $A(Q)$, energy centre, $E_{c}$, and energy width $w$.  One of these is an inelastic DHO (DHO1), centred at a relatively large energy, near 0.4 meV, while the other is assumed to be quasi-elastic (DHO2), whose $E_c$ is constrained to be 0.01~meV in all fits (Precise value of $E_c$ for DHO2 is immaterial as it is much smaller than its width). The elastic scattering region, $E = [-0.08, 0.08]$~meV, is excluded from these fits, as this region is dominated by temperature-independent incoherent elastic scattering, as well as elastic magnetic diffuse scattering for temperatures near and below $T_g$ (see next section).

Both DHO lineshapes individually satisfy $S(\textbf{Q},E) = \chi^{\prime\prime}(\textbf{Q},E) \times \frac{1}{1-e^{-E/K_{\mathrm{B}}T}}$ with $\chi^{\prime\prime}(\textbf{Q},E)$ an odd function of energy, and therefore both satisfy the detailed balance condition.

This form accounts well for the continuum of inelastic scattering at all temperatures.  Representative fits of this form of $S(Q,E)$ are shown in Fig.~\ref{Tempdep_fit} at four different $Q$ values from $Q=0.5$~$\angstrom^{-1}$ to 1.7~$\angstrom^{-1}$, and from well below $T_g \sim 2.5$~K (at $T=0.125$~K), to 2~K and 4~K. Temperature dependence of the parameters for DHO1 and DHO2 ($E_c$ is allowed to vary only for DHO1) at Q=1.3$\AA^{-1}$ are shown Fig.~\ref{Tempdep_parameters} (a) and (b), respectively.

In Fig.~\ref{Tempdep_fit}, one can see that the high energy DHO (DHO1) is well defined at low temperatures, and broadens out with increasing temperature, but even at $T=4$~K, a well defined inelastic peak remains. Its amplitude displays a modest increase roughly correlated with $T_g \sim 2.5$~K. Its energy softens appreciably at $\sim$ $T_g$, with a concomitant increase in its energy width, such that the two cross at $\sim$ 5 K.  $E_c$ of DHO2 (0.01 meV) is temperature independent by construction.  Its amplitude falls off appreciably at $T_g$ and beyond, while its width weakly decreases with increasing temperature beyond $T_g$.

We can also examine the low energy inelastic neutron scattering data at low temperatures, with a high temperature $T=100$~K dataset subtracted. This is useful as it allows diffuse elastic magnetic scattering to be separated from temperature-independent, incoherent elastic scattering, and it also shows where the low temperature inelastic spectral weight exceeds that at high temperature. This latter feature needs to be examined in the context that magnetic inelastic scattering is still present at high temperatures, although typically softened in energy and more diffuse in $Q$ than at lower temperatures.

Figure~\ref{tempsubtraction} shows a set of four such temperature subtractions: $T = 0.125$~K - $T = 100$~K, $T = 1.5$~K - $T = 100$~K, $T = 4$~K - $T = 100$~K, and $T = 10$~K - $T = 100$~K as a function of $Q$ and energy, with net intensity shown on a logarithmic color scale. Any negative net intensity (high temperature intensity $>$ low temperature intensity) is shown white.  The elastic magnetic diffuse scattering near and below $T_g$ is easily identified as the most intense temperature dependent scattering in this system.  One may be tempted to associate the low energy negative net scattering in the temperature-subtracted data of Fig.~\ref{tempsubtraction}~(a), between $\sim$~0.08 meV and $\sim$~0.2~meV, as the presence of a gapped excitation spectrum at low temperatures, but our earlier DHO analysis of the inelastic magnetic scattering \textit{without} a temperature subtraction shows no such evidence for a gap - we simply identify this energy regime as the regime where the dynamic spectral weight at high temperatures is greater than that at low temperatures, likely due to the overall softening of the magnetic inelastic spectral weight as a function of temperature and the effect of the Bose thermal population factor.

\section{Analysis of Elastic Neutron Scattering} 

The temperature subtraction shown in Fig.~\ref{tempsubtraction} clearly demonstrates that, in addition to the low energy quasi-elastic scattering (accounted for by DHO2 in Fig.~\ref{Tempdep_fit}), there exists a \textit{static} component due to the development of frozen spin correlations around $T_g$ and below.  Unlike the low energy inelastic scattering, the resulting temperature-subtracted elastic data is not affected by the temperature dependence of the Bose population factor. 

\begin{figure}[]
  \centering
  \includegraphics[width=0.4 \textwidth]{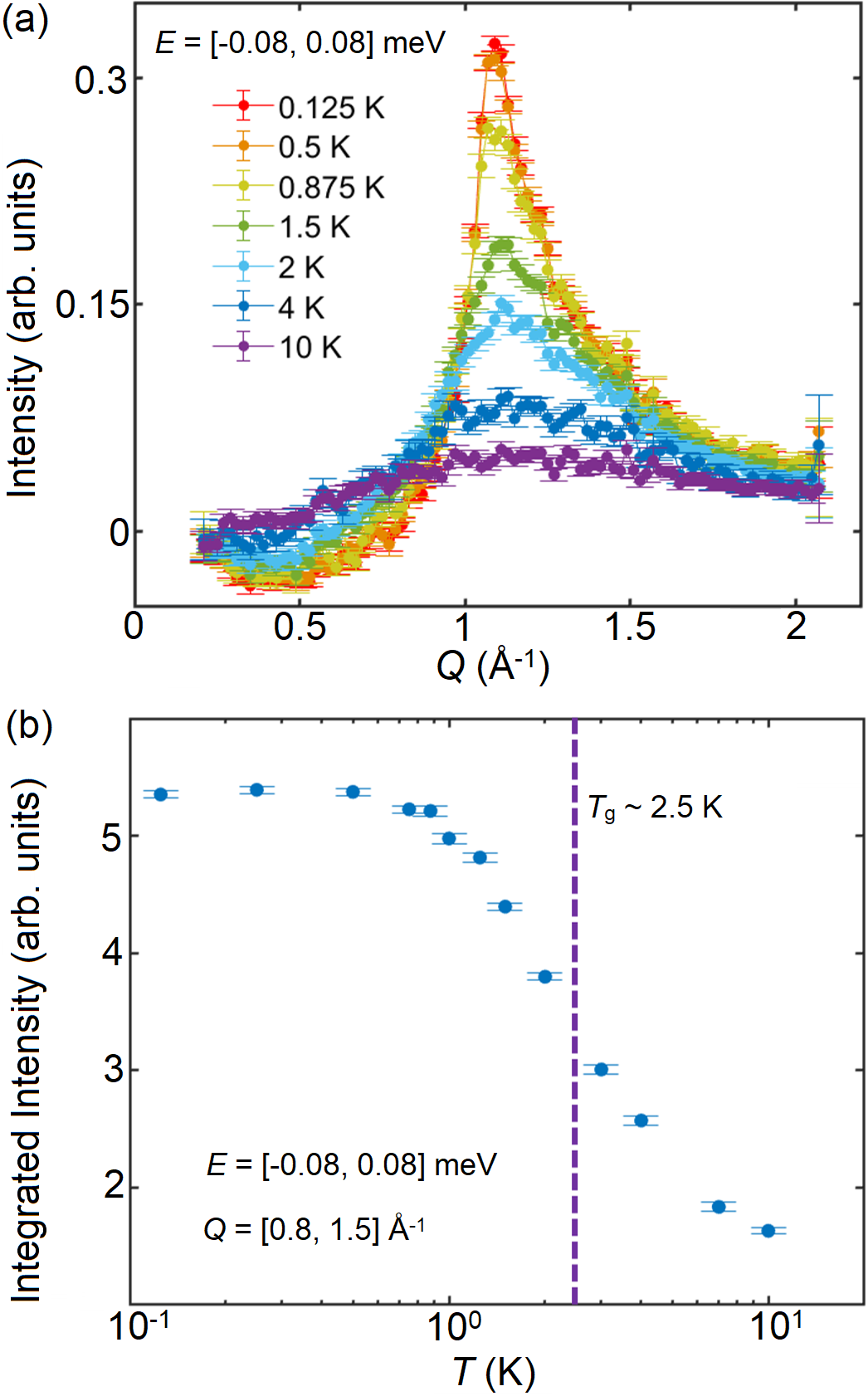}
  \caption{(a) The $Q$-dependence of the temperature-subtracted elastic neutron scattering signal measured from powder ErMgGaO$_4$ in this work, for temperatures between $T = 0.125$~K and $T = 10$~K, with a $T = 100$~K dataset subtracted in each case and an energy integration range of $E = [-0.08, 0.08]$~meV. (b) The temperature dependence of the integrated intensity corresponding to the temperature-subtracted data shown in (a) for a $Q$ integration range of $[0.8, 1.5]$~$\angstrom^{-1}$. The dashed line in (b) shows an approximate inflection point of the integrated intensity as a function of temperature, at $T_g \sim 2.5$~K.}
   \label{elastic} 
\end{figure}

Figure~\ref{elastic}~(a) shows the temperature evolution of the ErMgGaO$_4$ elastic neutron scattering data, integrated over the energy range $E = [-0.08, 0.08]$~meV, for temperatures between $T=0.125$~K and 10~K with a $T = 100$~K dataset subtracted in each case. The observed net elastic intensity peaks in Fig.~\ref{elastic} are relatively sharp in $Q$ at low temperatures but skewed out to higher Q, similar to what is expected from powder averaged two dimensional rods of scattering that are described by Warren lineshapes~\cite{Clark2019}.  These features clearly broaden out in $Q$ with increasing temperatures. The integrated intensities from 0.8 to 1.5~$\angstrom^{-1}$ are plotted in Fig.~\ref{elastic}~(b) as a function of temperature. The inflection point of this integrated intensity curve exactly coincides with the glass transition temperature, $T_g \sim 2.5$~K, from the magnetization data (Fig.~\ref{Figure2}). The temperature dependence of this net elastic intensity flattens out at below $T= 0.8$~K, and maintains this level down to the lowest temperatures measured, $T=0.125$~K. \\

\subsection{Warren Line Shape Analysis of Diffuse Elastic Scattering}

As discussed above, the $Q$-dependence of the ErMgGaO$_4$ elastic scattering ($E = [-0.08, 0.08]$~meV) resembles a Warren lineshape, which originates from the powder average of two dimensional rods of scattering.  The Bragg-like features in a Warren lineshape possess a sharp, low-Q onset, but the high-Q tail extends out so as to give a markedly asymmetric lineshape, at least when the two dimensional correlation length significantly exceeds that in the third dimension (which is assumed to be zero in the Warren lineshape).

\begin{figure*}[]
  \centering
  \includegraphics[width=0.75 \textwidth]{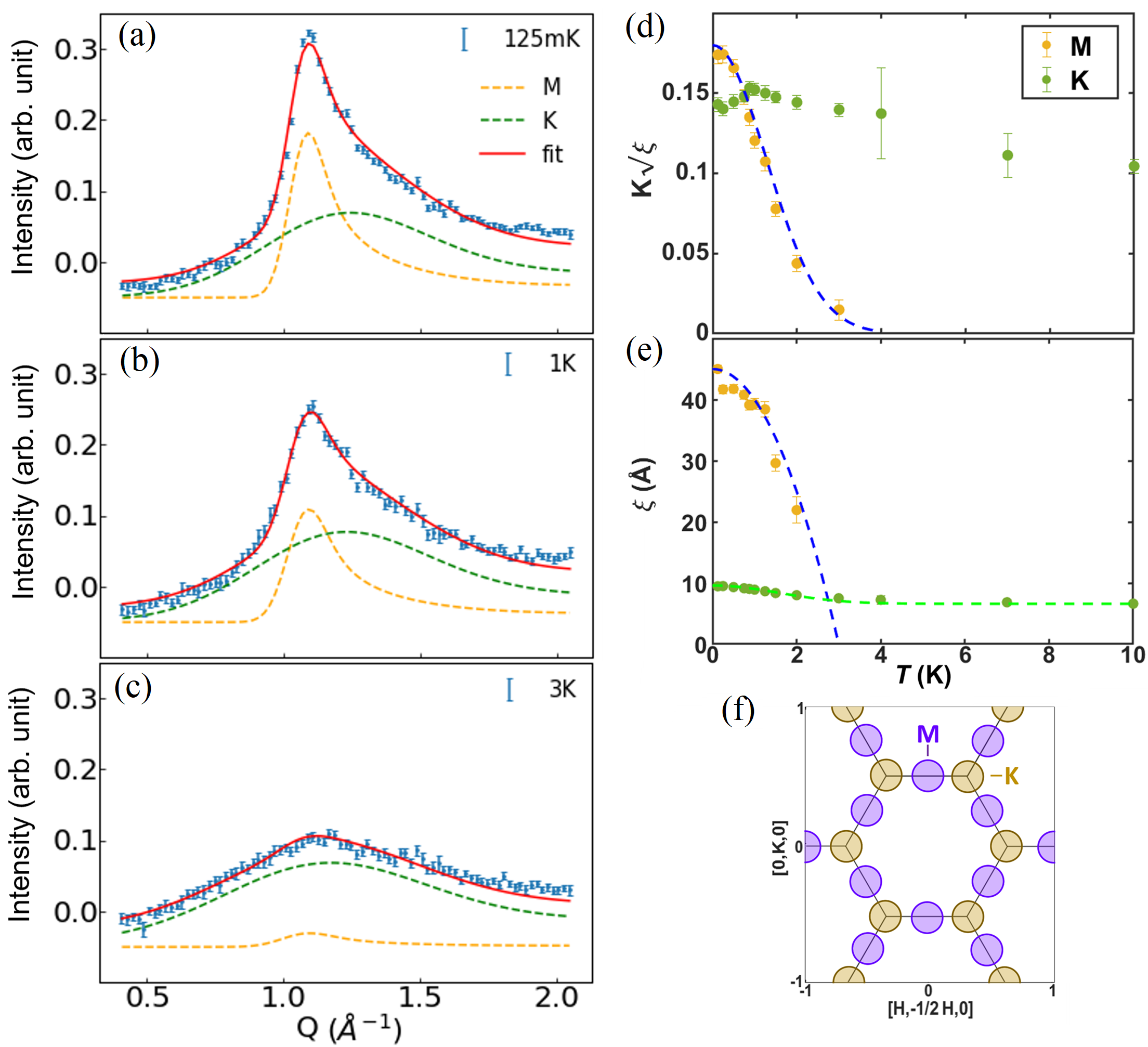}
  \caption{Warren lineshape fits to the elastic neutron scattering signal measured from ErMgGaO$_4$ at (a)~$T = 0.125$~K, (b)~$T = 1$~K, and (c)~$T = 3$~K. In each case, the fit results in red consist of the the sum of the Warren lineshape centered at the M point (yellow dashed line) and the Warren lineshape centered at the K point (green dashed line), along with an additional temperature independent background of -0.03 units. We show both the M-point and K-point fits with an offset of -0.05 units for visual clarity. (d)~The temperature dependence of the integrated intensity for the Warren lineshapes centred at the M and K points, estimated using $K \sqrt{\xi}$ where $K$ and $\xi$ are as defined for each Warren lineshape in Eq.~\eqref{eq:4}. (e) The temperature dependence of the correlation length $\xi$ (see Eq.~\eqref{eq:4}) for the Warren lineshapes centred at the M and K points. The dashed lines in (d) and (e) are guides to the eye. (f)~The $(H,K-\frac{H}{2},0)$ plane of reciprocal space showing the locations of the M (purple) and K (gold) points.}
   \label{warrenanalysis} 
\end{figure*}  

While ErMgGaO$_4$'s elastic diffuse scattering is asymmetric and resembles a simple Warren lineshape, this data cannot be reasonably described by a single Warren lineshape.  Instead at least two Warren-lineshapes are required for a high quality fit at low temperatures, and this is what is shown as the solid and dashed lines in Fig.~\ref{warrenanalysis}.  The presence of two Warren lineshapes makes physical sense if two different two-dimensional magnetic structures compete and co-exist, which occurs naturally within a $J_1-J_2$ model as shown by the zero temperature phase diagram in Fig.~\ref{phase diagram}.  Competing two-dimensional stripy and 120$^\circ$ structures would then give rise to co-existing Warren lineshapes centred at the M point and K points on the hexagonal zone boundaries in reciprocal space [Fig.~\ref{warrenanalysis}~(f)] which, upon powder-averaging, are peaked at $Q_0 = 1.06$~$\angstrom^{-1}$ and 1.25~$\angstrom^{-1}$, respectively.

This motivates us to fit the elastic diffuse scattering to two Warren line shapes, centered at $|Q|$'s corresponding to the M and K points in reciprocal space, respectively (Fig.~\ref{warrenanalysis}). Each individual Warren line shape equation \citep{Clark2019} can be written as:\\
\begin{align}\label{eq:4}	
I(Q) = & Km\frac{F^2_{hk}[1-2(\frac{\lambda Q}{4 \pi})^2 + 2(\frac{\lambda Q}{4 \pi})^4]}{(\frac{\lambda Q}{4 \pi})^{3/2}}  \nonumber  \nonumber \\ & \times (\frac{\xi}{\gamma\sqrt{\pi}})^{1/2} F(a)[f(Q)]^2 ~~,
\end{align}
where, 
\begin{align}\label{eq:5}	
a = \frac{\xi\sqrt{\pi}}{2\pi}(Q – Q_0)~~, 
\end{align}
and, 
\begin{align}\label{eq:6}
F(a) = \int_{0}^{10}e^{-(x^2-a^2)^2}dx ~~.
\end{align}

Here, $K$ is a scaling constant, $m$ is the multiplicity of the reflection, $F_{hk}$ is the two dimensional structure factor for the spin array, $\lambda$ is the neutron wavelength, $\xi$ is the two-dimensional spin-spin correlation length, $Q_0$ is the centre of the peak, and $f(Q)$ is the magnetic form factor for Er$^{3+}$. 

 At low temperature, the elastic diffuse scattering intensity is the largest at the M point, but significant diffuse intensity is required at both the M and K points in order to adequately describe the elastic diffuse scattering. With increasing temperature, the M point scattering drops while that at the K point is maintained. Fig.~\ref{warrenanalysis}~(d,e) shows the total integrated intensity of the Warren line shape at both high symmetry points and the two dimensional correlation lengths relevant to each structure as a function of temperature.  Specifically, Fig.~\ref{warrenanalysis}~(d) shows the product of $K$ and $\sqrt\xi$ from Eq.~\eqref{eq:4}, which is a proxy for the total intensity, and Fig.~\ref{warrenanalysis}~(e) shows the correlation length $\xi$ for the Warren lineshapes centered on the M and K points. 
 
 Figure~\ref{warrenanalysis}~(d) shows that the elastic intensity associated with the M point, due to the frozen short-ranged collinear stripy N\'eel state, drops rapidly with increasing temperature and is zero by $\sim$~4~K. In contrast, the K point elastic intensity, due to the frozen short-ranged non-collinear 120$^\circ$ N\'eel state, is maintained over this temperature range, falling off only weakly to $T=10$~K. In a similar fashion, Figure~\ref{warrenanalysis}~(e) shows that the correlation length $\xi$ associated with the M point elastic scattering decays significantly with increasing temperature while that of the K point elastic scattering remains relatively stable. 
 
The temperature dependence of these fit parameters is consistent with ErMgGaO$_4$ hosting a coexistence of frozen collinear stripy short-ranged order along with frozen non-collinear 120$^\circ$ short-ranged order in its ground state, with the frozen stripy correlations dominating at temperatures below $T_g \sim 2.5$~K. With increasing temperature, the stripy correlations rapidly dissolve, leaving non-collinear 120$^\circ$ short range order for temperatures above $T_g$. \\

\subsection{Monte Carlo Modeling of \texorpdfstring{$J_1$}~-\texorpdfstring{$J_2$}~-\texorpdfstring{$\Delta$}~ Model and Co-existing Stripy and 120\texorpdfstring{$^\circ$}~ Order}

\begin{figure*}[]
\centering
  \includegraphics[width=0.96 \textwidth]{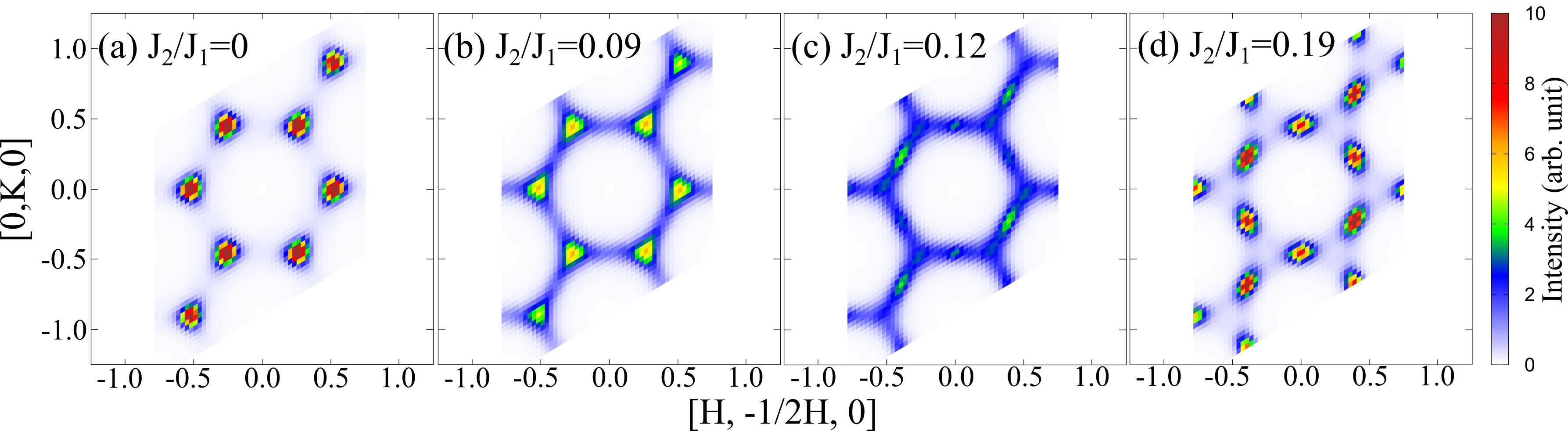}
\caption{Classical Monte Carlo calculations for the equal-time structure factor in the $(H,K,0)$ plane of reciprocal space according to Eq.~\eqref{eq:2} with $J_{\pm\pm} = J_{z\pm} = 0$ and $\Delta = 0.4$. The temperature employed was $J_1$/10 in all cases.}

   \label{montecarlo} 
\end{figure*}  

We have carried out classical Monte Carlo simulations of Heisenberg spins interacting via the $J_1$-$J_2$-$\Delta$ model on a triangular lattice so as to better understand the nature of the low temperature ground state in ErMgGaO$_4$.  

These calculations have been carried out at low temperatures using Eq.~\eqref{eq:2}, with $\Delta$ varying from 0 to 1, and the bond-dependent terms set to zero, $J_{z\pm} = J_{\pm\pm} = 0$. The Fourier transform of these Monte Carlo calculations, giving $S(Q)$, are shown as a function of temperature in Fig.~\ref{montecarlo} for $\Delta$=0.4. Fig.~\ref{montecarlo} (a-d), shows $S(Q)$, for $J_2/J_1$=0, 0.09, 0.12, and 0.19, respectively.  Fig.~\ref{montecarlo}~(c) which shows $S(Q)$ for $J_2/J_1=0.12$ and $\Delta=0.4$, is close to the classical expectation appropriate to the best fit parameters we have estimated for ErMgGaO$_4$, as indicated in Fig. \ref{phase diagram}.

At the lowest temperatures, the classical calculations show either long range K-point (non-collinear 120$^\circ$ N\'eel state) order for $J_2/J_1=0$, or quasi-long-ranged M point (collinear stripy state) order for $J_2/J_1=0.19$, as is expected in the theoretically established $J_1$-$J_2$ phase diagram (Fig.~\ref{phase diagram}). In the case of $J_2/J_1=0$, the K point correlations dominate over M point correlations at all temperatures.  However, this is not the case for $J_2/J_1=0.12$, the case closest to that appropriate to ErMgGaO$_4$, where M point correlations co-exist with short range correlations at all zone boundary positions even at the lowest temperatures. In particular, $S(Q)$ is evenly distributed along zone boundaries at high temperatures. 

These classical calculations are consistent with our placement of ErMgGaO$_4$ near the classical phase boundary between a stripy collinear N\'eel phase and a non-collinear 120$^\circ$ N\'eel phase.

\section{Conclusion}

We have synthesized and characterized approximately phase pure powder samples of ErMgGaO$_4$, wherein pseudospin-1/2 degrees of freedom associated with Er$^{3+}$ are arranged on a stacked triangular lattice.  This triangular lattice quantum antiferromagnet is the sister compound of the well studied YbMgGaO$_4$, and both display disordered Mg$^{2+}$/Ga$^{3+}$ bilayers between the rare-earth triangular layers~\cite{Li2015, Paddison2017, Cava2018}. Magnetic susceptibility measurements at low temperature in our powder samples show a FC/ZFC bifurcation indicative of a spin glass transition in ErMgGaO$_4$ at $T_g \sim 2.5$~K, roughly 1/6 of its $\Theta_{CW}$ ($\sim -14$~K).  

High energy neutron spectroscopy reveals well-defined, although not resolution limited, CEF transitions with a bandwidth of $\sim$ 80 meV.  This spectrum of CEF states originates within the $J=15/2$ multiplet appropriate for Er$^{3+}$, and shows a surprising low energy first-excited CEF state $\sim$~3~meV above the ground state. This is much lower than expected from earlier single-ion calculations~\cite{Cai2020}, and is likely to give rise to interesting virtual crystal field transitions, which are known to alter the expected terms in the spin Hamiltonian~\cite{Rau2016, RauReview2019}.  The best fit solution to the CEF Hamiltonian predicts clear XY anisotropy (0$\leq  \Delta \leq $1), distinct from the Ising expectations of the relevant point charge model.

The low energy spin excitation spectrum shows what appears to be a continuum of scattering not unlike the expectations of spinons in quantum disordered ground states. This low energy magnetic spectral weight has a bandwidth of $\sim$~0.8~meV, and a strong elastic component freezes out below $T_g \sim 2.5$~K. The low-energy magnetic spectral weight can be phenomenologically described as a sum of two DHOs, the lower energy one of which is overdamped at all temperatures and describes the quasi-elastic scattering. The higher energy DHO is well defined in energy at temperatures similar to and below $T_g$, and it defines the $\sim$~0.8~meV bandwidth of the low energy inelastic scattering. 

A LWST description of the powder-averaged low energy magnetic spectral weight can account for many of the experimentally-observed features, using a $J_1-J_2-\Delta$ model on a triangular lattice with $\frac{J_2}{J_1}=0.13 \pm 0.03$ and $\Delta=0.4 \pm 0.1$, which places ErMgGaO$_4$ on the $J_1-J_2-\Delta$ phase diagram for the triangular lattice, within the stripy N\'eel phase, immediately above the predicted spin liquid phase. The remaining discrepancies between the observations and the calculations can be reasonably ascribed to an inadequacy of LWST due to enhanced quantum fluctuations close to the phase boundary, and/or the known inherent disorder present in ErMgGaO$_4$ and the related materials within this family. The substantial disorder also explains the significant spin-glass or frozen nature of the ground state in ErMgGaO$_4$ .

The elastic ($-0.08$~meV$< E <+0.08$~meV) component to the magnetic spectral weight is diffuse in $Q$ at all temperatures, but with an asymmetric form characteristic of strong quasi-two-dimensional correlations. This was described as the sum of two Warren lineshapes, which represent co-existing and competing two-dimensional stripy collinear N\'eel and 120$^\circ$ non-collinear N\'eel states. This modeling shows that $T_g \sim 2.5$~K is associated with the development of the M-point intensity, characteristic of stripy collinear short range order, while the 120$^\circ$ non-collinear short range order (manifested as elastic intensity at the K-point), and is present at all temperatures.
The fact that the stripy 2D correlations win out at low temperatures is consistent with our placement of ErMgGaO$_4$ within the theoretical $J_1-J_2-\Delta$ phase diagram.


\begin{acknowledgments}

We acknowledge illuminating discussions with Alexandre L. Chernyshev.  This work was supported by the Natural Sciences and Engineering Research Council of Canada (NSERC). We greatly appreciate the technical support from Qiang Chen and Marek~Kiela at the Brockhouse Institute for Materials Research, McMaster University. The synthesis at Princeton University was supported by the US Department of Energy under the C2QA program. A portion of this research used resources at the Spallation Neutron Source, a DOE Office of Science User Facility operated by the Oak Ridge National Laboratory. The beamtime on the SEQUOIA instrument was allocated on Proposal No. IPTS-28967.1.\\
\end{acknowledgments}

\section*{Appendix A: Further Details on Crystal Electric Field Analysis}
\subsection{Fitting Constant-$|Q|$ Cuts}
\begin{figure*}[]
\centering
  \includegraphics[width=0.96 \textwidth]{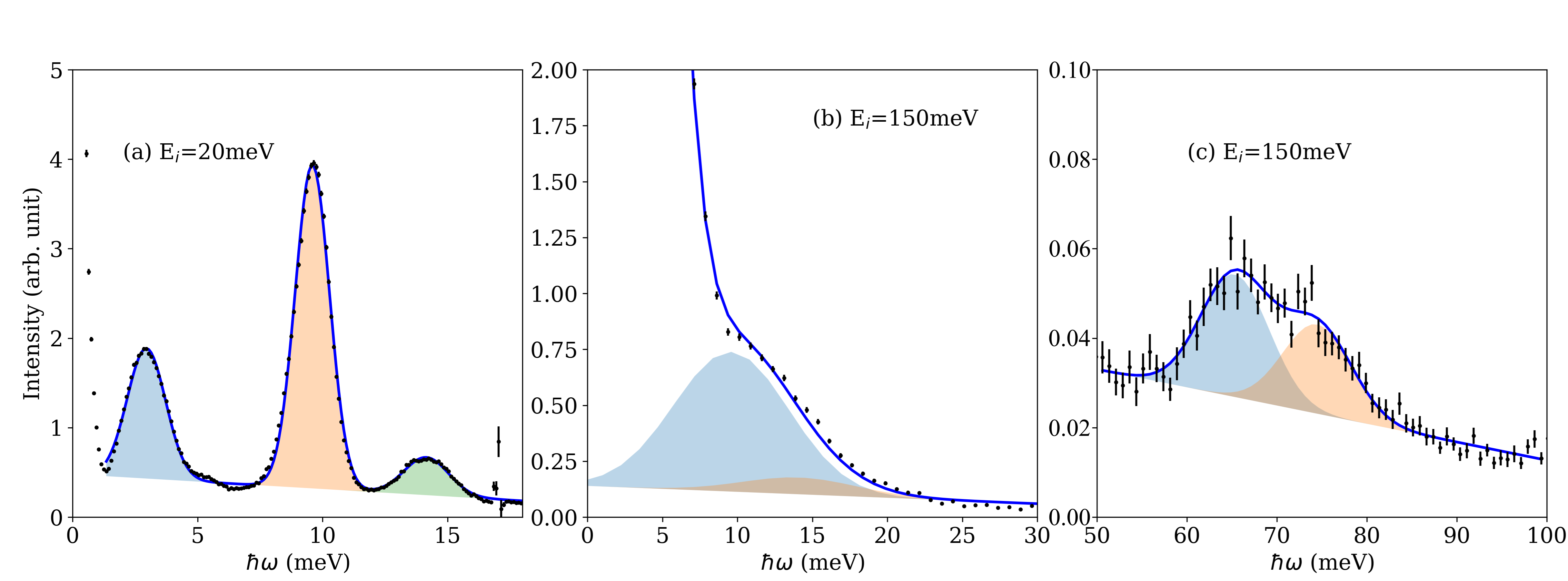}
\caption{Fitting constant-$|Q|$ cut of the (a) $\mathrm{E_i}=20~$meV and (b,c) $\mathrm{E_i}=150~$meV data to a sum of Gaussians as described in the Appendix.} 
   \label{fitting} 
\end{figure*}  
As shown in Fig.~\ref{fitting}~(a), the constant-$|Q|$ cut of the $\mathrm{E_i}=20~$meV data was fit to a sum of 3 Gaussians on top of a linear background. Only data outside the elastic line are considered in the fit. Since the CEF peaks are clearly broader than the instrumental resolutions ($\Delta=0.53~$meV) at $\mathrm{E_i}=20~$meV, their widths are allowed to vary in the fit. The energies and peak intensities of the three CEF transitions thus obtained are given in Fig.~\ref{TableI}, with a FWHM of 1.8(3)~meV [Transition $\# 1$], 1.6(2)~meV [Transition $\# 2$] and 2.1(3)~meV [Transition $\# 3$], respectively.

Since the low and high energy CEF transitions have very different intensities, fitting to the $\mathrm{E_i}=150~$meV data was carried out separately for $\mathrm{E}<50~$meV and $\mathrm{E}>50~$meV as shown in Fig.~\ref{fitting}~(b) and (c), respectively. A shoulder centered at $\sim10$~meV is clearly resolved next to the elastic peak in Fig.~\ref{fitting}~(b), which we attribute to the overlap of transitions $\# 2$ and $\# 3$ (transition $\# 1$ is buried under the elastic peak). The data with $\mathrm{E}<50~$meV is therefore fit to a sum of two Gaussians [shaded in Fig.~\ref{fitting}~(b)] whose energies and relative intensities are constrained to the values obtained from the $\mathrm{E_i}=20~$meV data. Since intrinsic widths of the CEF transitions are $\sim2$~meV $\ll \Delta E\sim9~$meV at $\mathrm{E_i}=150~$meV, widths of the two Gaussians are fixed to $\Delta E$. The resulting best fit is shown as blue solid line in Fig.~\ref{fitting}~(b). Assuming the same widths (=$\Delta E$) for the high energy CEF transitions, data in Fig.~\ref{fitting}~(c) are best fit with two overlapping Gaussians centered at $\sim 65~$meV and $\sim 75~$meV, respectively.

\subsection{Generalized Simulated Annealing}
Energies and intensities of the CEF transitions given in Table~\ref{TableI} are used to constrain the CEF parameters in Eq.~\eqref{eq:1} using the \texttt{dual\_annealing} algorithm (as implemented in \texttt{scipy.optimize}) with a cost function given by Eq.~\eqref{cost}. The \texttt{CrystalField} package in \texttt{Mantid} was used to evaluate the energies and intensities of the CEF transitions for a given set of $\{B_m^n\}$. Under a generic CEF with $D_{3d}$ point group, one expects 8 Kramers' doublets from the $J=\frac{15}{2}$ multiplet, and hence 7 transitions at T=0. The observation of only 5 transitions in our experiment might therefore be either due to the presence of two or more transitions that are too close to be resolved, or to the small intensities of some transitions. Given the large widths of the CEF peaks observed experimentally, we used a fairly generous energy tolerance level, $\tau$, of 0.5~meV and 5~meV for transitions with $\mathrm{E}<20~$meV and $\mathrm{E}>20~$meV, respectively. In other words, two transitions, i and j, with an energy difference, $|\mathrm{E}_i -\mathrm{E}_j|<\tau$, are treated as a single transition with a total intensity of $I_1+I_2$. We then evaluate $\mathcal{C}$ (Eq.~\eqref{cost}) with the five most intense transitions while discarding the weaker ones. $\{B_m^n\}$ giving fewer than 5 transitions below $\mathrm{E}<100~$meV are rejected by assigning a large $\mathcal{C}=2.5$ (corresponding to an average $\left|\tfrac{X^\mathrm{exp}-X^\mathrm{theo}}{\sigma}\right|\sim300$).  

To explore only physically relevant parameter space, we considered two sets of trial CEF parameters: those computed directly from a point charge model,
${B_m^n}_\mathrm{,pc}=\{-3.44 \times 10^{-1}, -5.32 \times 10^{-4},-1.22 \times 10^{-2}, 2.42 \times 10^{-6}, 1.32 \times 10^{-5},0\}$\citep{Cai2020}
and those estimated from the experimentally determined CEF parameters for YbMgGaO$_6$\citep{Li2017} using the relation $B_m^n (\mathrm{Er^{3+}})=\frac{\theta_n^\mathrm{Er^{3+}}\left\langle r^n\right\rangle_{\mathrm{Er^{3+}}}}{\theta_n^\mathrm{Yb^{3+}}\left\langle r^n\right\rangle_{\mathrm{Yb^{3+}}}}B_m^n (\mathrm{Yb^{3+}})$ \citep{Hutchings1964}:
$B_{m\mathrm{,est}}^n=\{-1.13 \times 10^{-2}, 
-5.31 \times 10^{-5}, 
2.76 \times 10^{-2}, 
-1.60 \times 10^{-5}, 
-4.16 \times 10^{-4}, 
3.20 \times 10^{-4}
\}$. The range of $B_m^n$ explored in the search is taken to be $\pm \max\{\left|{B_m^n}_\mathrm{,pc}\right|\times10, \left|{B_m^n}_\mathrm{,est}\right|\times10, 0.001\}$.

\subsection{CEF Parameters}
Searches carried out with random initial guesses all converge to $\mathcal{C}\lesssim 0.3$ or $\left|\tfrac{X^\mathrm{exp}-X^\mathrm{theo}}{\sigma}\right|\lesssim 1$, implying a good fit to the low temperature INS data. Two clearly superior sets of CEF parameters (out of 20 searches) are shown in Table~\ref{parameters} with their corresponding $\mathcal{C}$-values. Given the under-constrained nature of the problem (fewer transitions are observed experimentally than are theoretically allowed), as well as the complex $\mathcal{C}$-landscape with many local minima and discontinuities, we do not claim our search is exhaustive and other solutions likely exist. 

\begin{table*}[h]
\centering
\begin{tabular}{c c c c c c c c}
\hline
Set & $B_{2}^{0}$ & $B_{4}^{0}$ & $B_{4}^{3}$ & $B_{6}^{0}$ & $B_{6}^{3}$ & $B_{6}^{6}$ & $\mathcal{C}$ \\
\hline
A & $5.53 \times 10^{-1}$ & $2.67 \times 10^{-3}$ & $3.83 \times 10^{-2}$ & $3.27 \times 10^{-5}$ & $-6.59 \times 10^{-5}$ & $7.82 \times 10^{-5}$ & 0.19 \\
B & $1.40 \times 10^{-1}$ & $-2.14\times 10^{-3}$ & $-1.30\times 10^{-2}$ & $-9.36 \times 10^{-6}$ & $2.14 \times 10^{-4}$ & $-1.02 \times 10^{-4}$ & 0.16\\

\hline
\end{tabular}
\caption{Two best sets of CEF parameters with their corresponding cost function values $\mathcal{C}$. Predictions using set A are presented in the main text}
\label{parameters}
\end{table*}

\begin{figure*}[]
\centering
  \includegraphics[width=0.96 \textwidth]{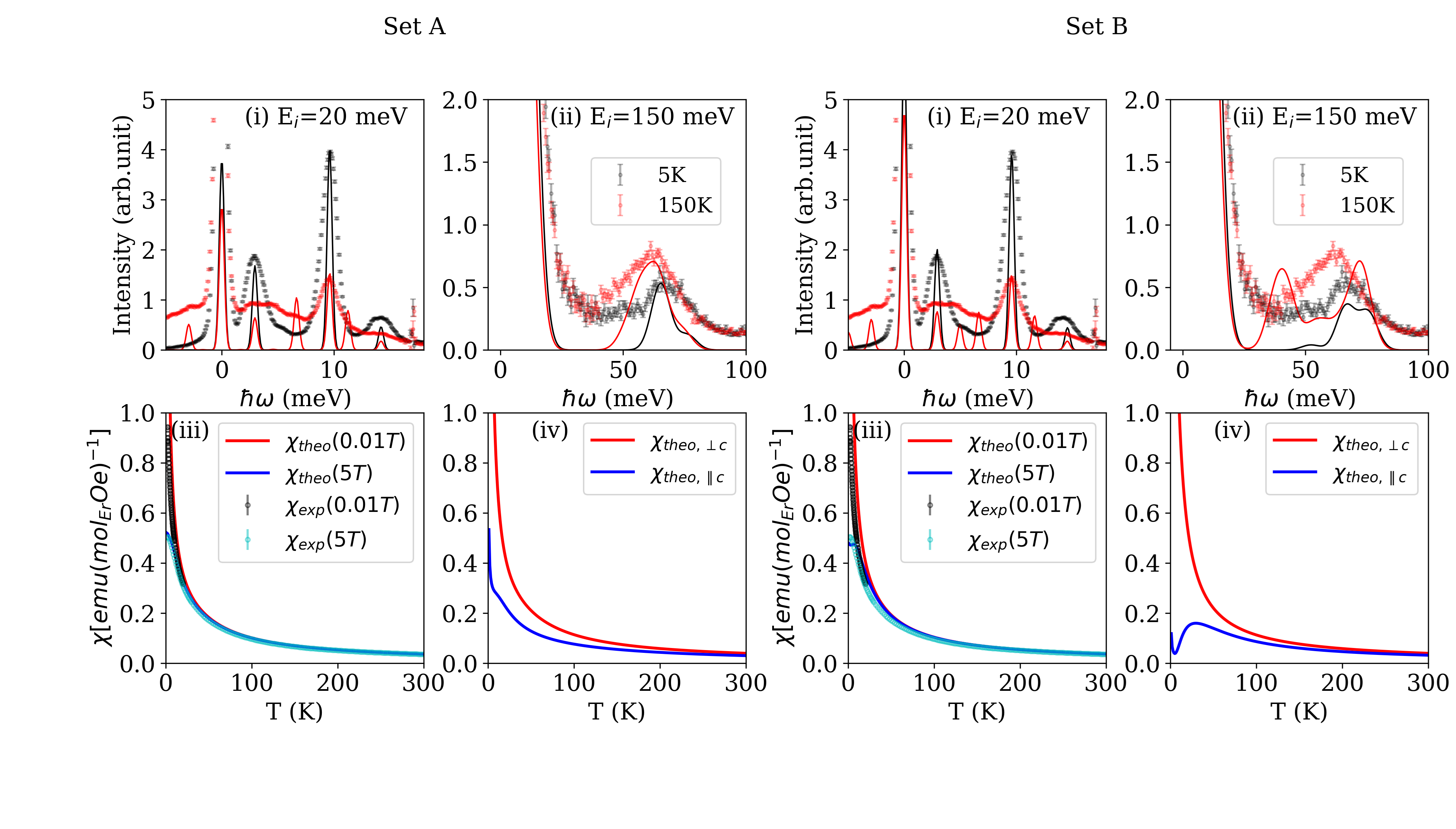}
\caption{Comparison between theory and experiments for the two sets of CEF parameters shown in Table~\ref{parameters}. (i) and (ii) show constant $|Q|$ cuts at 5~K (black open circle) and 150~K (red open circle) along with the corresponding theoretical calculations (solid lines) for (i) $\mathrm{E_i}=20~$meV and (i) $\mathrm{E_i}=150~$meV. The same scaling factor is applied to the theoretical curves at different temperatures, which have been convolved with the instrumental resolutions for each $\mathrm{E}_i$. (iii) Experimental and theoretical magnetic susceptibilities at different fields (same as Fig.~\ref{Figure5} in the main text). (iv) Predicted single-crystal magnetic susceptibility for different field directions. } 
   \label{expvstheory} 
\end{figure*}

In Fig.~\ref{expvstheory}, we compare the theory with different experiments for the two sets of CEF parameters shown in Table~\ref{parameters}. As shown by the black data points in (i), both parameters reproduce the energies and intensities of the CEF peaks in the $\mathrm{E_i}=20$~meV data very well, with set B yielding better slightly agreement with the $\mathrm{E_i}=$150~meV data- hence a slightly lower $\mathcal{C}$. To provide further constraint for the CEF parameters, we also show the 150~K data in (i) and (ii) for the two $\mathrm{E}_i$'s as red circles. The three lowest CEF peaks are all broader and weaker at high temperature but show little change in peak positions, whereas the peak at $\sim70~$meV softens considerably and \textit{gains} intensity. As shown clearly in Fig.~\ref{expvstheory}, set A (presented in the main text) predicts an increase in intensity and softening for the high-energy CEF peak qualitatively consistent with the experiment, while set B predict very different behaviours.

As shown in (iii), the calculated magnetic susceptibilities using both parameters agree well with the experiment at high temperatures but deviate at low temperatures, especially for the 0.01~T data. The agreement at low temperatures is slightly better for set A. In (iv), we simulated the temperature-dependent single-crystal magnetic susceptibilities. Both parameters show an easy-plane magnetic anisotropy at all temperatures albeit with very different temperature dependence, the latter of which can be used to further constrain the CEF parameters when good quality single crystal samples become available. Spectroscopically, combining other techniques (e.g. Raman scattering and optical conductivity \citep{CEFCsErSe2} that follow different selection rules) with neutron scattering might allow a unique determination of the CEF parameters in this compound. 

\begin{figure*}[]
\centering
  \includegraphics[width=0.96 \textwidth]{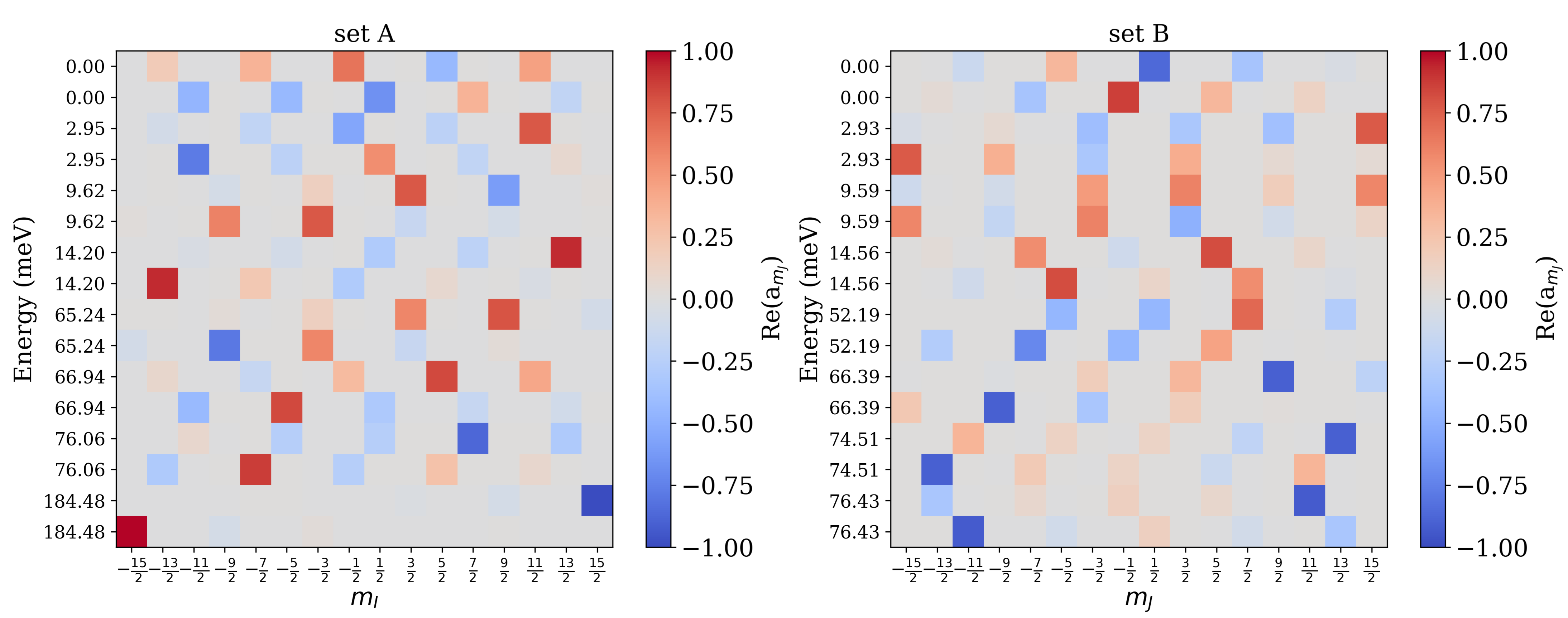}
\caption{The eigenvalues and eigen-wavefunctions (expressed in the $m_J$-bases) for the two sets of parameters in Table~\ref{parameters}. The real part of the amplitudes, $\mathrm{Re}(a_{m_J})$ [defined by $\left| i\right\rangle =\sum_{m_J} a_{m_J}\left| m_J\right\rangle$], for the $i^\mathrm{th}$ eigenstate are shown on a color scale along the $i^\mathrm{th}$-row of each panel. $m_J$ values of the bases functions are given on the horizontal axis, and the energy of the eigenstates are shown on the vertical axis.} 
   \label{states} 
\end{figure*}  

The eigenvalues and eigen-wavefunctions associated with the CEF parameters in Table~\ref{parameters} are shown in Fig.~\ref{states}. Despite the drastically different $B_m^n$-values, an important observation
is that the ground doublet for all parameter sets are comprised of only the $\{\left|\pm\tfrac{11}{2}\right\rangle,\;
\left|\pm\tfrac{5}{2}\right\rangle,\;
\left|\pm\tfrac{1}{2}\right\rangle,\;
\left|\pm\tfrac{7}{2}\right\rangle,\;
\left|\pm\tfrac{13}{2}\right\rangle\}$-bases with the $\left|\pm\tfrac{1}{2}\right\rangle$ component having the largest weight. Remarkably, this is true for \textit{all} parameters with $\mathcal{C}\lesssim 0.3$, suggesting a robust XY-like single-ion anisotropy associated with the GS doublet in Er$^{3+}$. However, the wave-functions of the excited CEF levels vary significantly across different parameter sets, giving rise to the drastically different temperature-dependence observed in magnetic susceptibilities and INS spectra in Fig.~\ref{expvstheory}. 

\subsection{Effects of Disorder}
\begin{figure*}[]
\centering
  \includegraphics[width=0.96 \textwidth]{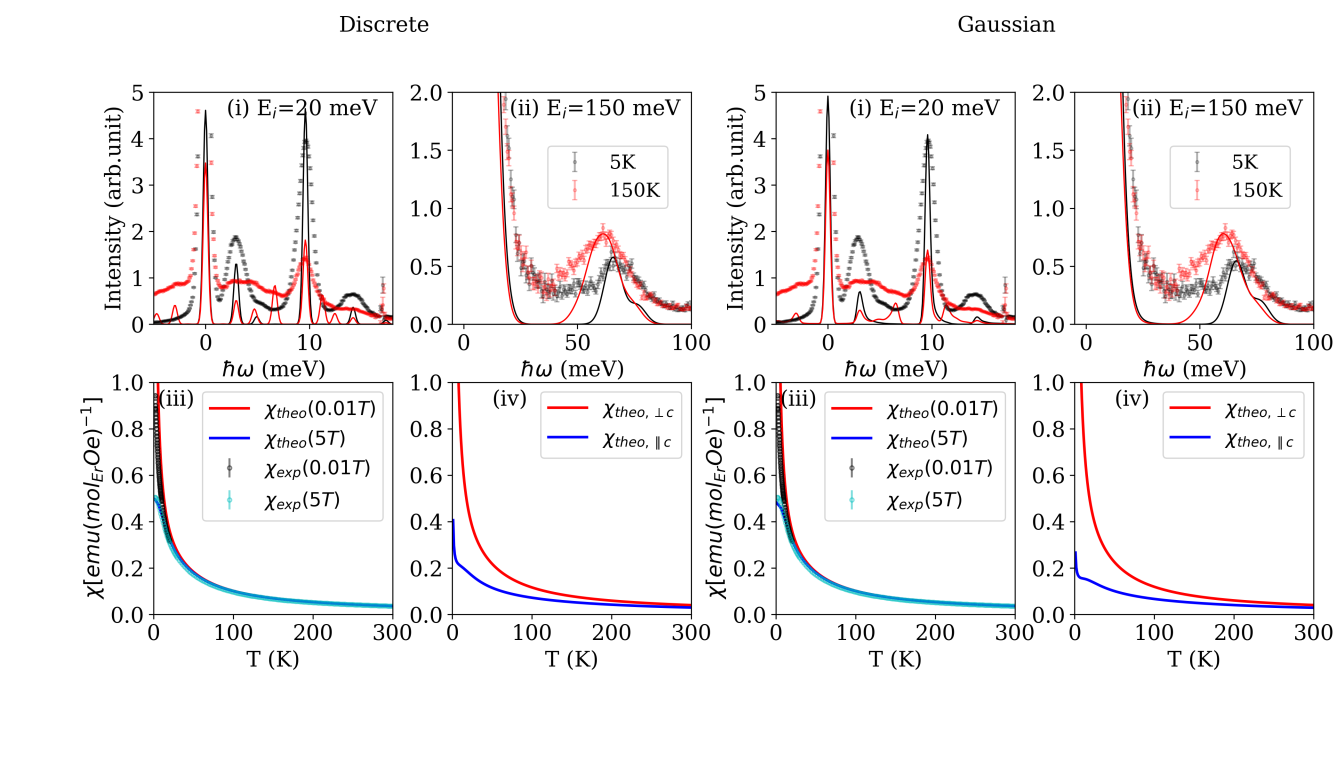}
\caption{Comparison between theory and experiments for the two distributions of Er$^{3+}$ positions along $z$, or $P(z)$, in a model with structural disorder. See Appendix for details. The panels are shown in the same order as Fig.~\ref{expvstheory}.} 
   \label{disorder} 
\end{figure*}

The presence of Er$^{3+}$ positional disorder has been noted in Ref.~\citep{Cava2018}. Instead of forcing Er$^{3+}$ to be at the high symmetry [0,0,0] position, a better fit to the single-crystal X-ray diffraction data was achieved by displacing Er$^{3+}$ away from the center of the octahedra by $z=\pm d_0\approx0.1\AA$ within a split-site model. We explore the effect of Er$^{3+}$ positional disorder on CEF spectra in this section. Even with a $z\neq 0$, the CEF Hamiltonian is constrained to take the form of Eq.~\eqref{eq:1} by the $C_{3v}$ symmetry (instead of $D_{3d}$ for $z=0$). Although the CEF parameters obtained from the point-charge model, ${B_m^n}_\mathrm{,pc}$ \citep{Cai2020}, deviate significantly from the best fit (set A in Table~\ref{parameters}, or ${B_m^n}_\mathrm{, A}$), we use it to estimate the CEF parameters at a given $z$ by the following relation\citep{Li2017}: ${B_m^n} (z)=\frac{{B_m^n}_\mathrm{,pc}(z)}{{B_m^n}_\mathrm{,pc}(0)} {B_m^n}_\mathrm{, A}$. We then model the INS spectra and the magnetic properties by averaging over different $z$-values following a given distribution $P(z)$. As shown in Fig.~\ref{disorder}, we considered both (a) a discrete model where $z$ takes on discrete values of $\{0, \pm d_0\}$, and (b) Gaussian model where $P(z)$ is a Gaussian with a full-width-at-half-maximum of $2d_0\approx0.2\AA$. Clearly, the discrete model provides a better description of the INS spectra at both temperatures. Notably, the CEF parameter corresponding to $z=\pm d_0$ gives rise to the experimentally observed weak peak at $\sim 5~$meV. Within this model, the intensity ratio of peaks at 2.95~meV and 5~meV, $\sim 10$ from the data in Fig.~\ref{Figure4}, constrains the occupancy of the high-symmetry $z=0$ site and $z=\pm d_0$ sites to be $\sim$0.7 and $\sim$0.3 (used in Fig.~\ref{disorder} and Fig.~\ref{Figure4}), respectively. Further \textit{ab initio} calculation is required to provide a microscopic explanation for the value of $d_0\approx0.1\AA$, as well as the preference for a discrete rather than a continuous distribution of $z$.

%

\end{document}